\documentclass[aip,graphicx]{revtex4-2}

\usepackage{amsmath}
\usepackage{graphicx} 
\usepackage{subfig} 
\usepackage{amssymb}
\usepackage{trfsigns}
\usepackage{color}
\usepackage{algorithm}
\usepackage[usenames,dvipsnames]{xcolor}
\usepackage{dcolumn}
\usepackage{mathtools}
\usepackage{tikz}
\usepackage{appendix}
\usepackage{amsfonts}
\usepackage{amsmath}
\usepackage{amssymb}
\usepackage{amsthm}
\usepackage{float}
\usepackage{subfig}
\usepackage{dsfont}

\floatname{algorithm}{Table}
\definecolor{MyGreen}{rgb}{0.45,0.60,0.09}

\DeclareRobustCommand{\frac}[3][0pt]{%
  {\begingroup\hspace{#1}#2\hspace{#1}\endgroup\over\hspace{#1}#3\hspace{#1}}}

\begin{document}


\title{The shape of an axisymmetric meniscus in a static liquid pool: effective implementation of the Euler transformation}  



\author{Nastaran Naghshineh}
\email[corresponding author: ]{nxncad@rit.edu}
\affiliation{School of Mathematics and Statistics, Rochester Institute of Technology, Rochester, NY, 14623, USA}
\affiliation{Department of Mathematics and Sciences, Rochester Institute of Technology-Dubai, Dubai, 341055, UAE}

\author{W. Cade Reinberger}
\affiliation{School of Mathematics and Statistics, Rochester Institute of Technology, Rochester, NY, 14623, USA}

\author{Nathaniel S. Barlow}
\affiliation{School of Mathematics and Statistics, Rochester Institute of Technology, Rochester, NY, 14623, USA}

\author{Mohamed A. Samaha}
\affiliation{Department of Mechanical and Industrial Engineering, Rochester Institute of Technology-Dubai, Dubai, 341055, UAE}
\affiliation{School of Mathematics and Statistics, Rochester Institute of Technology, Rochester, NY, 14623, USA}
        
\author{Steven J. Weinstein}
\affiliation{School of Mathematics and Statistics, Rochester Institute of Technology, Rochester, NY, 14623, USA}

\affiliation{Department of Chemical Engineering, Rochester Institute of Technology, 
Rochester, NY, 14623, USA}



\date{\today}
\begin{abstract}
{We examine the classical problem of the height of a static liquid interface that forms on the outside of a solid vertical cylinder in an unbounded stagnant pool exposed to air. Gravitational and surface tension forces compete to affect the interface shape as characterized by the Bond number. Here, we provide a convergent power series solution for interface shapes that rise above or fall below the horizontal pool as a function of contact angle and Bond number. We find that the power series solution expressed in terms of the radial distance from the wall is divergent, and thus rewrite the divergent series as a new power series expressed as powers of an Euler transformed variable; this series is modified to match the large distance asymptotic behavior of the meniscus. The Euler transformation maps non-physical singularities to locations that do not restrict series convergence in the physical domain, while the asymptotic modification increases the rate of convergence of the series overall. We demonstrate that when the divergent series coefficients are used to implement the Euler transformation, finite precision errors are incurred, even for a relatively small number of terms. To avoid such errors, the independent variable in the governing differential equation is changed to that of the Euler transform, and the power series is developed directly without using the divergent series. The resulting power series solution is validated by comparison with a numerical solution of the interface shape and the small Bond number matched asymptotic solution for the height of the interface along the cylinder developed by Lo (1983, J. Fluid Mech., 132, p.65-78). The convergent power series expansion has the ability to exceed the accuracy of the matched asymptotic solution for \textit{any} Bond number given enough terms, and the recursive nature of the solution makes it straightforward to implement.}
\end{abstract}
\maketitle 
\section{Background and Formulation} \label{sec:Intro}
The shape of a static interface (meniscus) formed outside a vertical cylinder is partially submerged in an infinite horizontal pool has been studied since the early 1900s. This problem was first examined in the literature by~\cite{Ferguson} who obtained solutions for cylinders having large radii. Since then, the problem has been solved numerically for wire coating and other various applications~\citep{WhiteTallmadge, HuhScriven}.~\cite{James} and later~\cite{Lo} used the method of matched asymptotic expansions~\citep{VanDyke} to predict explicit expressions for the height of the meniscus along the cylindrical wall in the limit of small Bond number, $B$, as defined in~(\ref{eq:DimensionlessVars}) below. More recent extensions of this configuration is found in coating of optical fiber sensors and microfluids~\citep{FiberCoating,MicroFluid} in liquid pools of finite extent. 

The problem examined here is configured as follows. An infinite horizontal pool of liquid with density, $\rho$, is subject to gravity, $g$, and is in contact with air of negligible density. A cylindrical wall of radius, $R$, is placed vertically in the pool and the liquid intersects the wall with a contact angle, $\theta\in(0,\pi)$ (measured through the liquid). The air--liquid interface has surface tension, $\sigma$, and its height above (or below) the static pool, located at $z=0$, is parameterized as $z=H(r)$ where $r$ is the radial distance from the axis of symmetry of the cylinder. For this configuration, the Young-Laplace equation couples with the hydrostatic pressure field to yield the following dimensionless equation and boundary conditions for $\bar{r} \in [1,\infty)$:
\begin{subequations}
\begin{equation}
    \frac{d}{d\bar{r}}\left[\frac{\bar{r}\frac{d\bar{H}}{d\bar{r}}} {\left[1+(\frac{d\bar{H}}{d\bar{r}})^{2}\right]^{1/2}}\right] - B\bar{r}\bar{H} = 0, 
    \label{eq:FirstScaledODE}
\end{equation}
\begin{equation}
    \frac{d\bar{H}}{d\bar{r}} = -\cot\theta~\textrm{at} \ \bar{r}=1, 
    \label{eq:SlopeBoundaryCondition}
\end{equation}
\begin{equation}
    \bar{H}\rightarrow 0~\textrm{as} \ \bar{r}\rightarrow\infty,
    \label{eq:StaticPoolBC}
\end{equation}
where 
\begin{equation}
    \bar{r}= \frac{r}{R}~,~\bar{H}= \frac{H}{R}~,~B = \frac{\rho g R^{2}}{\sigma}.
    \label{eq:DimensionlessVars}
\end{equation}
    \label{eq:FirstScaledODESystem}
\end{subequations}
In~(\ref{eq:DimensionlessVars}), the overbars denote dimensionless variables, and $B$ is the Bond number, which provides a ratio of characteristic scales for gravitational stress, $\rho g R$, and surface tension stress, $\sigma / R$, that compete to deform the interface. 

The nonlinear system~(\ref{eq:FirstScaledODESystem}), may be solved approximately on a finite domain by shooting, collocation, or other numerical methods. The intention of this work is to provide, for the first time, an analytical solution for the interface shape governed by~(\ref{eq:FirstScaledODESystem}) over the entire semi-infinite domain, which is expressed here as a power series. Although the standard power series solution to~(\ref{eq:FirstScaledODESystem}) diverges (as shown herein), we combine two resummation techniques to overcome this difficulty and to construct a single convergent expansion. Recently,~\cite{FlatWallSakiadisPaper,NonNewtonian} constructed convergent series solutions for the  boundary layer along a moving wall (the Sakiadis Boundary Layer and its non-Newtonian counterpart) by judiciously choosing an independent variable motivated by its asymptotic behavior at large distances from the wall. In this work, we employ a similar approach to construct a convergent series solution that is asymptotically consistent at both ends of the domain. We combine this asymptotic consistency with the well-known Euler transformation~\citep{VanDyke} to arrive at our convergent power series solution. The relatively simple structure of this problem enables us to examine the efficacy of asymptotically consistent power series resummation methods for solving nonlinear ODEs introduced by~\cite{Barlow:2017}. 

We validate our power series solution by comparing its results with a numerical solution, as well as the $B \rightarrow 0$ matched asymptotic solution for the height of the meniscus at the cylinder by~\cite{Lo}. As a general note here, asymptotic series are often divergent and are thus limited by optimal truncation constraints. Additionally, the effort the effort to determine higher order corrections can be significant~\citep{VanDyke}. On the other hand, asymptotic methods are particularly efficient at describing limiting regimes with just a few terms and are a staple of fluid mechanics and other disciplines. The advantage of the power series method used here is that no overlap region is required as in matched asymptotics, and, by virtue of being a convergent power series series, our solution may be refined to within machine precision. 

As stated above, the Euler transformation is embedded in our power series result. It is known that when the coefficients of the original divergent power series are used to construct the Euler coefficients, finite precision errors inherent to the original series coefficients become magnified~\citep{Scraton}---this restricts solution accuracy beyond a certain number of series terms. However, we show in this paper that the computational issue may be circumvented by transforming the radial variable in the system~(\ref{eq:FirstScaledODESystem}) to the Euler transform variable and obtaining the power series solution to the transformed system directly. This result is important as this deficiency in the Euler transformation has been observed in other problems of mathematical physics (see, for example, \citet{Boyd1999}).

The paper is organized as follows. In section~\ref{sec:DivergentSection}, the power series solution to the system~(\ref{eq:FirstScaledODESystem}) is obtained and is found to be divergent.  The series is transformed via resummation in section~\ref{sec:ConvergentSection} such that it matches the large distance asymptotic behaviour and the influence of convergence-limiting singularities are mapped outside of the transformed domain via an Euler transformation; this results in a convergent representation of the solution as a power series in the Euler transformation variable. Section~\ref{sec:ConvergentResults} provides a comparison of the convergent power series and the numerical solution. Section~\ref{sec:HeightPrediction_ODETransformed} provides an algorithm, which uses the convergent series representation and Newton's method to predict the height of the interface at the cylinder wall. In section~\ref{sec:AsymptoticVsSeries}, the results are compared with the first and second order matched asymptotic solutions of~\cite{Lo}. Concluding remarks of this study are provided in section~\ref{sec:Conclusion}. In Appendix~\ref{sec:MatchedAsympSol}, the first and second order matched asymptotic solutions developed by~\cite{Lo} are rewritten for reference in the current notation, and in doing so, we correct two typos in that work. Formulae used to manipulate power series are provided in Appendix~\ref{sec:series}. Appendix~\ref{sec:SmallSlope} provides a useful solution, valid for all Bond numbers, for configurations where the value of the contact angle, $\theta$, lies near $\pi / 2$. Appendices~\ref{sec:Appendix_CoeffComputation} and~\ref{sec:Appendix_EuelerizedBessel} provide the details for determining the Euler coefficients by means of a variable substitution in the original governing ODE in system~(\ref{eq:FirstScaledODESystem}). 
\section{Divergent Power Series Solution} \label{sec:DivergentSection}
We now consider the solution of~(\ref{eq:FirstScaledODESystem}) via power series. To begin, we make the transformation
\begin{subequations}
\begin{equation}
    \tilde{r}=\bar{r} - 1,
\end{equation}
such that $\tilde{r}=0$ corresponds to the location where the meniscus meets  cylinder wall, and write $\bar{H}(\bar{r}) = \bar{H}(\tilde{r}+1) = \bar{h}(\tilde{r})$ in equation system~(\ref{eq:FirstScaledODESystem}) to obtain
\begin{equation}
    (\tilde{r}+1) \bar{h}'' = B(\tilde{r}+1) \bar{h}\left[ (\bar{h}')^{2}+1\right]^{3/2}- (\bar{h}')^{3} - \bar{h}',
    \label{eq:RearrangedODE}
\end{equation}
\begin{equation}
    \bar{h}' = -\cot\theta~\textrm{at} \ \tilde{r}=0,
    \label{eq:TransformedSlopeBC}
\end{equation}
\begin{equation}
     \bar{h}\rightarrow 0~\textrm{as} \ \tilde{r}\rightarrow\infty.
    \label{eq:TransformedStaticPoolBC}
\end{equation}
\label{eq:TransformedSystem}
\end{subequations}
\noindent 
\ignorespacesafterend
In~(\ref{eq:TransformedSystem}), note that we have utilized primes to denote derivatives of $\bar{h}$ with respect to $\tilde{r}$, and $\tilde{r}\in [0, \infty)$.

The standard power series solution of~(\ref{eq:TransformedSystem}) is found by first assuming the form
\begin{subequations}
\begin{equation}
    \bar{h}(\tilde{r})= \sum_{n=0}^\infty a_{n}~\tilde{r}^{n},~~|\tilde{r}|< \tilde{r}_s(B,\theta),
    \label{eq:AssumedSolu}
\end{equation}
where $\tilde{r}_s(B,\theta)$ is the yet undetermined radius of convergence as function of $B$ and $\theta$. Equation~(\ref{eq:AssumedSolu}) is then substituted into~(\ref{eq:RearrangedODE}) and---after applying JCP Miller's formula and Cauchy's product rule (see Appendix~\ref{sec:series})---terms of like powers are equated to obtain
\begin{equation}
    a_{n+2} = \displaystyle\frac{Bf_{n}-c_{n}-(n+1)a_{n+1}-n(n+1)a_{n+1}}{(n+1)(n+2)},
    \label{eq:a_n} 
\end{equation}
\begin{equation}
    f_{n>0} = e_{n-1}+e_{n},~f_{0} = e_{0},~c_{n}=\displaystyle\sum_{j=0}^n(n-j+1)b_{j} a_{n-j+1},
\end{equation}
\begin{equation}
    e_{n}= \displaystyle\sum_{j=0}^n (a_{j})(d_{n-j}),~d_{0}=
    \displaystyle\tilde{b}_{0}^{3/2},~d_{n>0}=\frac{1}{n\tilde{b}_{0}}\sum_{j=1}^n \left(\frac{5}{2}j-n\right)\tilde{b}_{j} d_{n-j},
\end{equation}
\begin{equation}
    \tilde{b}_{0} = 1 + b_{0},~\tilde{b}_{n>0} = b_{n>0},~ b_{n}=
    \displaystyle\sum_{j=0}^n(j+1)(n-j+1)a_{j+1} a_{n-j+1},
\end{equation}
with 
\begin{equation}
    a_0 = \bar{h}(0),~\textrm{and} \ a_1 = - \cot(\theta). 
    \label{eq:a_0}
\end{equation}
\label{eq:DivPowerSeries}
\end{subequations}
\noindent \ignorespacesafterend
Note that in~(\ref{eq:a_0}), the value of $a_0$ is not specified. 

\begin{figure}[H]
  \centering
  \includegraphics[width=9cm]{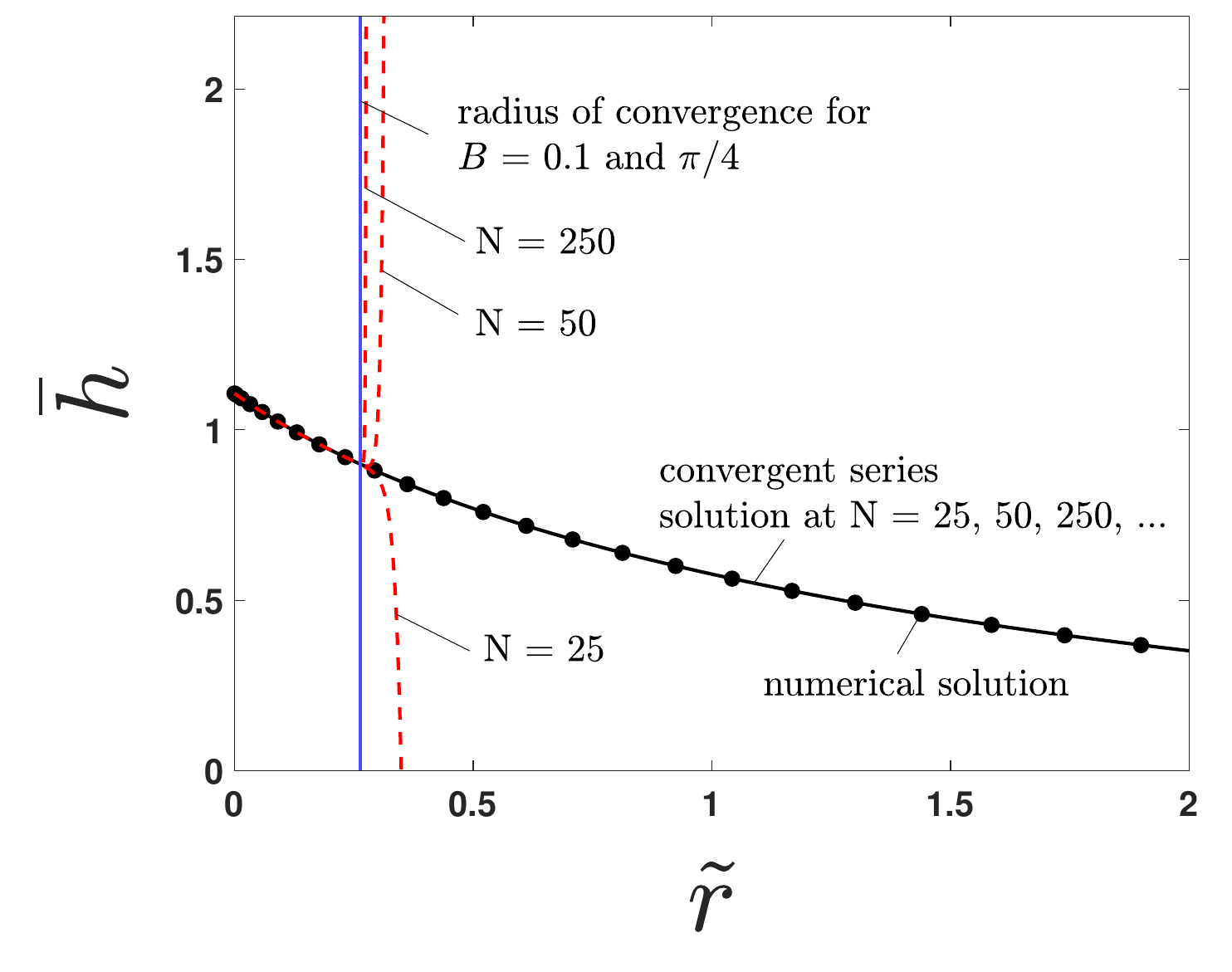}
  \caption{The solution to~(\ref{eq:TransformedSystem}) is shown for Bond number $B = 0.1$ and a contact angle of $\theta = \pi/4$. The $N$-term truncations of the divergent series solution~(\ref{eq:DivPowerSeries}) (\textcolor{red}{dashed curves}) and the convergent series solution~(\ref{eq:ConvPowerSeries_ODETransformed}) (solid curve) are compared against the numerical solution with $\mathcal{L}=60$~($\bullet$'s). The \textcolor{blue}{solid vertical line} shows the radius of convergence,~$\tilde{r}_s\approx 0.26$, computed from the Domb-Sykes plot.}
  \label{fig:DivSeriesSolu_45deg}
\end{figure}

\begin{figure}[H]
    \centering
    \includegraphics[width=9.5cm]{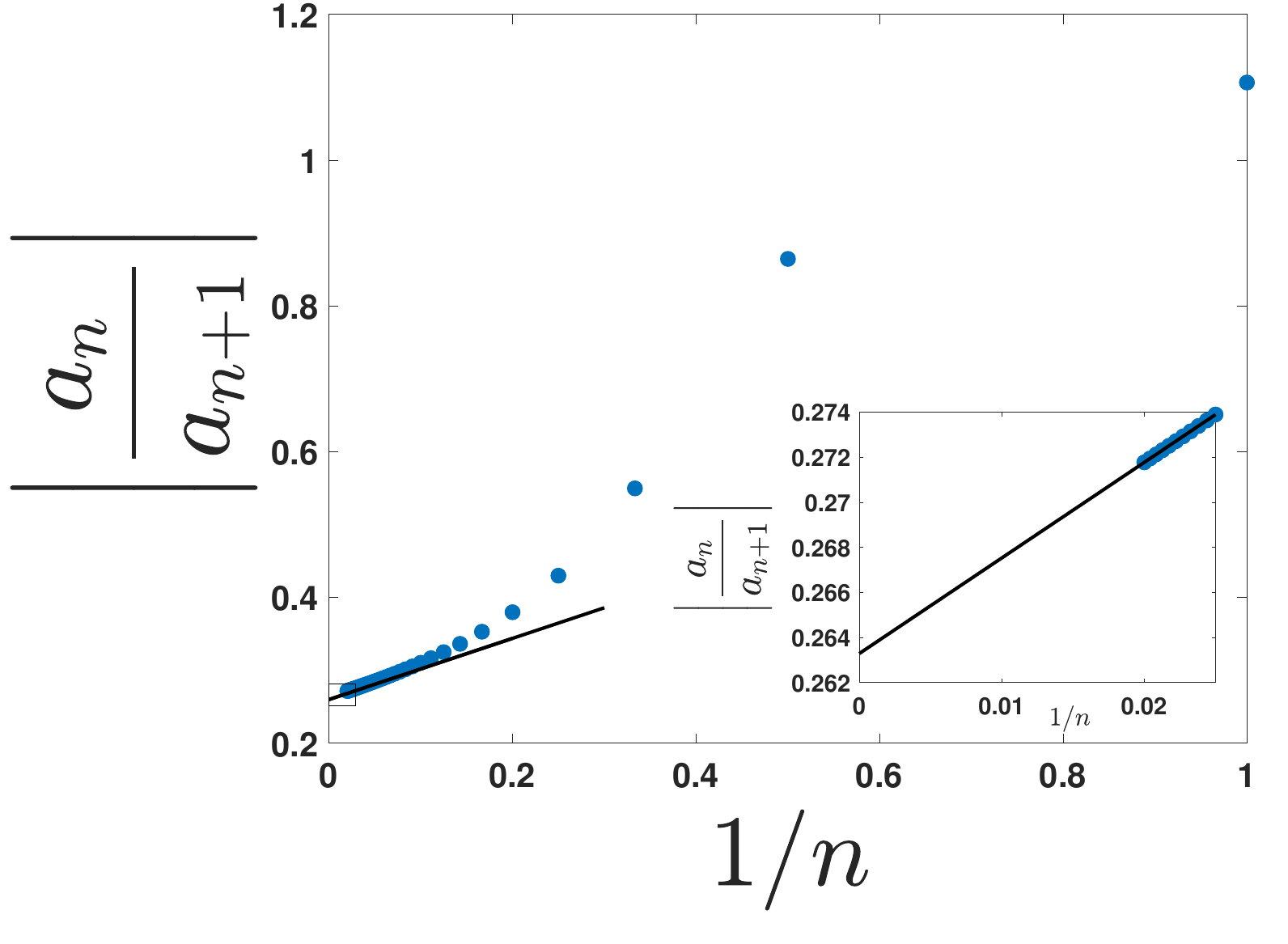}%
    \caption{Domb-Sykes plot for~(\ref{eq:DivPowerSeries}) with $B = 0.1$ and $\theta = \pi/4$ for $N=50$, indicating (via the ratio-test) a radius of convergence of $\tilde{r}_s\approx 0.26$. This is consistent with the divergent series behavior of~(\ref{eq:DivPowerSeries}) observed in figure~\ref{fig:DivSeriesSolu_45deg}.}%
    \label{fig:DivDombSykes_45deg}%
\end{figure}

At this point, we use the value of $a_0$ generated by a numerical solution to examine the behavior of the power series solution; in section~\ref{sec:HeightPrediction_ODETransformed} to follow, an algorithm to predict $a_0$ using the power series itself is provided. The numerical solution used here is a Chebyshev spectral method (the Chebfun package)~\citep{Chebyshev} that implements piecewise polynomial interpolation to solve the nonlinear differential equation as a boundary value problem on a \emph{finite} domain length $\mathcal{L}$; the length $\mathcal{L}$ is chosen so that doubling its size affects the finite domain solution below an infinity norm (taken between the $\mathcal{L}$ and $2\mathcal{L}$ solutions) of $O(10^{-13})$. In figure~\ref{fig:DivSeriesSolu_45deg}, the dashed curves show the $N$-term truncation of series~(\ref{eq:DivPowerSeries}) for $B=0.1$ and $\theta=\pi/4$. The power series solution~(\ref{eq:DivPowerSeries}) agrees with the numerical solution of~(\ref{eq:TransformedSystem}) for small $\tilde{r}$ but ultimately diverges at a finite radius of convergence (indicated as a solid vertical line in figure~\ref{fig:DivSeriesSolu_45deg}); the radius of convergence  $\tilde{r}_s\approx 0.26$, which is predicted using the Domb-Sykes plot in figure~\ref{fig:DivDombSykes_45deg} (the limit as $n \to \infty$ provides the radius of convergence); the Domb-Sykes plot itself is the numerical implementation of the ratio test~\citep{VanDyke}. Similarly divergent results are obtained for other values of $\theta$ and $B$.


\section{Convergent Resummation}  \label{sec:ConvergentSection}
\subsection{Euler Transformation}
Our goal is to recast the divergent series~(\ref{eq:DivPowerSeries}) as another power series that converges uniformly over the entire semi-infinite domain. This requires a variable transformation that maps the convergence-limiting singularity to a new location to increase the radius of convergence.  To do so, we note that the coefficients of the divergent series expansion~(\ref{eq:DivPowerSeries}) alternate in sign.  This indicates (by a corollary to Pringsheim's theorem) that the singularity responsible for divergence lies along the negative real axis~\citep{Hinch}. As such, it is likely that application of the Euler transformation will induce convergence.  

\begin{figure}[H]%
    \centering
    \includegraphics[width=8cm]{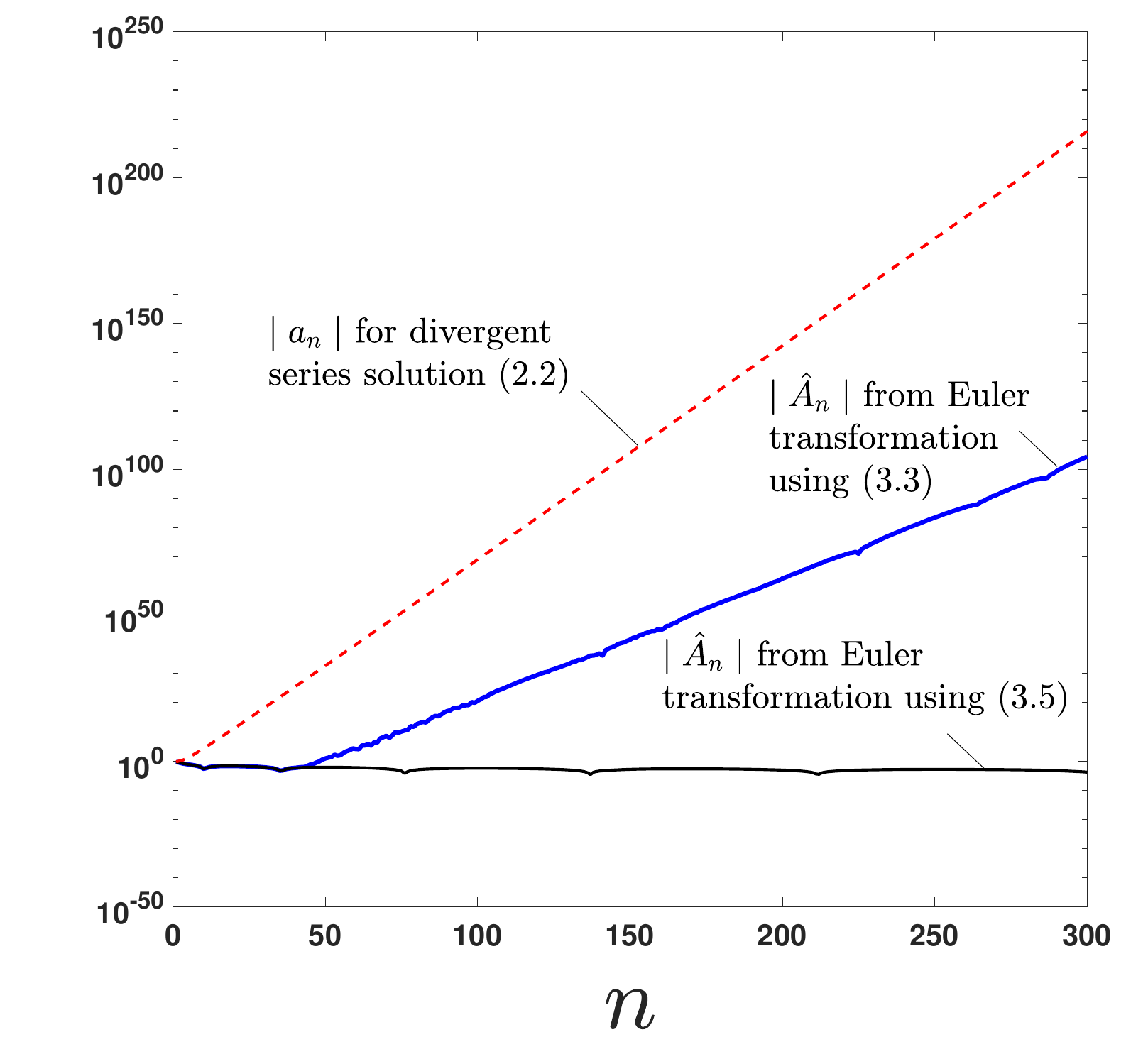}%
    \caption{Coefficients of the divergent series solution $a_n$ in~(\ref{eq:DivPowerSeries}) (\textcolor{red}{dashed curve}), Eulerized transformed coefficients $\hat{A}_n$ in~(\ref{eq:binomialEuler}) (\textcolor{blue}{solid curve}), and convergent series solution $\hat{A}_n$ in~(\ref{eq:ConvPowerSeries_ODETransformed}) (solid curve), plotted vs. $n$, for $\theta = \pi / 4$ and $B=0.1$ with $S=(2-\sqrt{2})/2$ calculated using~(\ref{eq:rocsmall}). All coefficients are computed in double precision.}%
    \label{fig:Coeff_vs_n}%
\end{figure}

The Euler transformation is defined as
\begin{subequations}
\begin{equation}
    \bar{h}(\tilde{r}) = g(\eta(\tilde{r})),
\end{equation}
with
\begin{equation}
  \displaystyle\eta = \frac{\tilde{r}}{\tilde{r}+S},
  \label{eq:Def_EulerTrans}
\end{equation}
\label{eq:EulerTrans} 
\end{subequations}
where $S>0$ is a parameter corresponding to the approximation of the radius of convergence of the diverging series~(\ref{eq:DivPowerSeries}).  Given a power series in $\tilde{r}$ with a negative real convergence limiting singularity located at $\tilde{r}= -\tilde{r}_s$, a power series in $\eta$, 
\begin{equation}
  g=\sum_{n=0}^\infty \hat{A}_n \eta^n,
  \label{eq:Def_EulerSeries}
\end{equation}
converges pointwise in $\eta$ for any positive real $S$ with $S <2\Tilde{r}_s$ (see Appendix~\ref{sec:ClosestSing}). The Euler transformation~(\ref{eq:EulerTrans}) maps the domain of the new expansion variable to be between $\eta = 0$ and $\eta = 1$, and maps the  singularity such that the transformed disc of convergence contains the entire physical domain, thus creating a convergent series. As shown in appendix~\ref{sec:Euler}, the power series coefficients of the transformed function $g$ about $\eta = 0$ can be written explicitly in terms of the original Taylor coefficients $a_n$ in~(\ref{eq:DivPowerSeries}) about $\Tilde{r}=0$ as
\begin{equation}
    \hat{A}_{n>0} = \sum_{m=1}^n  \binom{n-1}{m-1} a_m S^m,~\hat{A}_0=a_0.
    \label{eq:binomialEuler}
\end{equation}
The above relation is attractive in that it may be applied to any divergent alternating series with a finite radius of convergence, whether it originates as a solution to an ODE or not.  Thus, for purposes of generality, the Euler summation is often presented using an identity equivalent to~(\ref{eq:binomialEuler})~\citep{Hardy}. However, the implementation of Euler summation via~(\ref{eq:binomialEuler}) is computationally unstable to finite precision errors~\citep{Scraton}. This is shown in figure~\ref{fig:Coeff_vs_n}, where coefficients computed using~(\ref{eq:binomialEuler}) deviate from their true value beyond some finite $n$ value (here $ n \approx 50$). This error originates through the recurrence relation in~(\ref{eq:DivPowerSeries}) for the divergent series coefficients, since it incorporates the sums and differences of expressions involving exponentially growing series coefficients (see Figure~\ref{fig:Coeff_vs_n}). Since these coefficients are used in~(\ref{eq:binomialEuler}) to compute $\hat{A}_n$, the Euler summation coefficients incur this error. 

To avoid finite precision error in computing the Euler transformation~(\ref{eq:EulerTrans}), then, one must avoid using the original diverging series coefficients.  This is accomplished by applying the variable transformation~(\ref{eq:EulerTrans}) directly to the original ODE~(\ref{eq:RearrangedODE}) to obtain an ODE for $g(\eta)$.  The Taylor coefficients $\hat{A}_n$ are then found as the coefficients to the power series solution in $\eta$ to the transformed ODE. Upon applying the Euler transformation~(\ref{eq:EulerTrans}), the ODE (\ref{eq:TransformedSystem}) becomes
\begin{subequations}
\begin{equation} 
    g'' = \frac{2}{1-\eta}g' + \frac{((1-\eta)^4g'^2 + S^2)^{3/2} Bg}{(1-\eta)^4} - \frac{(1-\eta)^3g'^3}{S(\eta S+1-\eta)} -\frac{Sg'}{(1-\eta)(\eta S + 1 - \eta)},
    \label{eq:g_transformedODE}
\end{equation} 
\begin{equation}
    g'=-S \cot\theta ~\textrm{at} \ \eta = 0,
\end{equation}
\begin{equation}
    g \to 0 ~\textrm{as} \ \eta \to 1.
\end{equation}
\label{eq:g_transformedSystem}
\end{subequations}
In~(\ref{eq:g_transformedSystem}), the primes denote derivatives of $g$ with respect to $\eta$, and $\eta \in [0,1)$.  We next assume a solution to~(\ref{eq:g_transformedODE}) of the form
\begin{subequations}
\begin{equation}
    g(\eta) = \sum_{n=0}^\infty \hat{A}_{n}\eta^n,~~|\eta|< \eta_s(B,\theta).
    \label{eq:gseries}
\end{equation}
Using JCP Miller's formula~(\ref{eq:JCP}) and Cauchy's product formula~(\ref{eq:Cauchy}), a recurrence can be obtained (see Appendix~\ref{sec:Appendix_CoeffComputation}) for the Taylor coefficients $\hat{A}_n$ as 
\begin{equation}
    \hat{A}_{n+2} = \frac{S^2}{(n+1)(n+2)} \sum_{k=0}^n \binom{k+3}{3} w_{n-k},
\end{equation}
where
\begin{equation}
    w_n = f_n + \sum_{k=0}^n (q_k - t_k - u_k)r_{n-k}~~,~~f_n = \frac{2}{S^2}\sum_{k=0}^n (-1)^k\binom{3}{k}(n-k+1)\hat{A}_{n-k+1},
\end{equation}
\begin{equation}
    q_n = B \sum_{k=0}^n \ell_k p_{n-k}~~,~~\bar{\ell}_n = S + (1-S)\delta_{n,0}~~,~~\ell_n = \sum_{k=0}^n \bar{\ell}_k \hat{A}_{n-k}
\end{equation}
 \begin{equation}
      u_n = \frac{1}{S}\sum_{k=0}^n (-1)^k\binom{2}{k}(n-k+1)\hat{A}_{n-k+1} 
 \end{equation}
 \begin{equation}
     d_n = \sum_{k=0}^n u_k u_{n-k}~~,~~t_n = \sum_{k=0}^n d_k u_{n-k}~~,~~\bar{d}_n = d_n + \delta_{n,0} 
 \end{equation}
\begin{equation}
    r_n = (1-S)^n - (1-S)^{n-1}(1-\delta_{n,0}),
\end{equation}
\begin{equation}
 p_{n>0} = \frac{1}{n\bar{d}_0} \sum_{k=1}^n \left(\frac{5}{2}k-n \right)\bar{d}_kp_{n-k} ~~,~~ p_0 = \bar{d}_0^{3/2},
\end{equation}
with
\begin{equation}
     \hat{A}_0 = \bar{h}(\tilde{r}=0),~\textrm{and} ~~ \hat{A}_1 = \cot(\theta),
\end{equation}
where the Kronecker notation is used such that $\delta_{n,0}$ is 0 when $n\neq0$ and 1 when $n=0$. According to equation~(\ref{eq:gseries}) and the Euler transformation~(\ref{eq:EulerTrans}), the interface shape can be written in terms of physical domain coordinates as
\begin{equation}
\bar{h}(\tilde{r}) = \sum_{n=0}^\infty \hat{A}_n \left(\frac{\tilde{r}}{\tilde{r} + S}\right)^n.
\end{equation}
\label{eq:ConvPowerSeries_ODETransformed}
\end{subequations}
The result~(\ref{eq:ConvPowerSeries_ODETransformed}) is the desired uniformly convergent representation of the solution via the Euler transformation.  Returning to figure~\ref{fig:Coeff_vs_n}, a comparison is provided between the coefficients computed by transforming divergent coefficients according to~(\ref{eq:binomialEuler})  and those directly from the transformed differential equation according to~(\ref{eq:ConvPowerSeries_ODETransformed}). We observe that for $n<50$, coefficients obtained are identical.  However, for larger $n$ values, we see that the coefficients calculated from~(\ref{eq:binomialEuler}) begin to grow due to aforementioned finite precision error, and deviate from those obtained by~(\ref{eq:ConvPowerSeries_ODETransformed}). We note here that the solution~(\ref{eq:ConvPowerSeries_ODETransformed}) converges slowly, and that is evidenced by the slow decline in the magnitude of $\hat{A}_n$ as $n$ increases in figure~\ref{fig:Coeff_vs_n}. 

\subsection{Prefactor Inclusion}
Convergence of the power series solution~(\ref{eq:ConvPowerSeries_ODETransformed}) may be improved by incorporating the asymptotic behavior of the solution for $\Bar{h}(\Tilde{r})$ as $\Tilde{r} \to \infty$. We note that the power series solution~(\ref{eq:DivPowerSeries}) implicitly matches the correct asymptotic behavior as $\tilde{r}\rightarrow0$, and thus efficiently represents the solution near the cylinder wall. To guide the form of a convergent resummation of~(\ref{eq:DivPowerSeries}), the asymptotic behavior of $\bar{h}$ as $\tilde{r}\rightarrow \infty$ is first determined. The height constraint~(\ref{eq:TransformedStaticPoolBC}) necessarily implies that $\Bar{h}' \to 0$ as $\Tilde{r} \to \infty$. Thus, using the method of dominant balance on~(\ref{eq:RearrangedODE}) we neglect terms that are quadratic or higher in $\bar{h}'$ and obtain
\begin{equation}
    (\tilde{r}+1)^2 \bar{h}'' + (\tilde{r}+1)\bar{h}'-B(\tilde{r}+1)^2 \bar{h} = 0~\text{as} \ \tilde{r}\to\infty.
    \label{eq:AsympODE1}
\end{equation}
The solution of~(\ref{eq:AsympODE1}) is expressed in terms of the modified Bessel function (of the second kind) of zeroth order as
\begin{equation}
    \bar{h}(\tilde{r})\sim D~K_{0}\left(\sqrt{B}(\tilde{r}+1)\right)~\text{as} \ \tilde{r}\to\infty,
    \label{eq:prefactor}
\end{equation}
where $D$ is an undetermined constant that can only be found through consideration of finite $\Tilde{r}$ behavior. For $\theta$ near $\pi / 2$, note that the asymptotic result~(\ref{eq:prefactor}) is an excellent approximation for $\Bar{h}$ for all $\Tilde{r}$, and the boundary condition~(\ref{eq:TransformedSlopeBC}) may be applied to determine $D$ (see Appendix~\ref{sec:SmallSlope}).  For smaller angles, the power series~(\ref{eq:ConvPowerSeries_ODETransformed}) may be improved by combining the modified Bessel function~(\ref{eq:prefactor}) with the Euler transformation~(\ref{eq:Def_EulerTrans}) as
\begin{subequations}
\begin{equation}
   \bar{h}(\tilde{r}) = K_{0}\left(\sqrt{B}(\tilde{r}+1)\right) \sum_{n=0}^\infty \bar{A}_n~ \eta^n.
   \label{eq:BesselEuler_Initial}
\end{equation}
In~(\ref{eq:BesselEuler_Initial}), $\eta$ is the Euler transformation defined in~(\ref{eq:Def_EulerTrans}), and the estimate for the radius of convergence, $S$, is determined in what follows. Note that as $\tilde{r}\rightarrow\infty$, the series in~(\ref{eq:BesselEuler_Initial}) approaches a constant, preserving the asymptotic Bessel function behavior~(\ref{eq:prefactor}) in this limit. 

To solve for the Euler coefficients, $\bar{A}_n$, in~(\ref{eq:BesselEuler_Initial}), we use Cauchy's product formula (see Appendix~\ref{sec:Cauchy}), and the formula for the Taylor coefficients of $K_0(\sqrt{B}(\Tilde{r}+1))^{-1}$ in an expansion about $\eta=0$, relating $r$ and $\eta$ through~(\ref{eq:Def_EulerTrans}). This is done using the defining ODE for the modified Bessel function, and applying the same techniques used to compute the coefficients $\hat{A}_n$ in Appendix~\ref{sec:Appendix_CoeffComputation}; details are provided in Appendix~\ref{sec:Appendix_EuelerizedBessel} to obtain the coefficients 
\begin{equation}
    \bar{A}_n = \sum_{k=0}^n \zeta_k \hat{A}_{n-k},
\end{equation}
where
\begin{equation}
 \zeta_n = \frac{-1}{\bar{\zeta}_0}\sum_{k=1}^n \bar{\zeta}_k \zeta_{n-k} ~~,~~\zeta_0 = \bar{\zeta}_0^{-1} ~~,~~
    \bar{\zeta}_0 = K_0(\sqrt{B})~~,~~\bar{\zeta}_1 = -S\sqrt{B}K_1(\sqrt{B}),
\end{equation}
\begin{equation}
    \bar{\zeta}_{n+2} = \frac{1}{(n+1)(n+2)}\left[\sum_{k=0}^n\left\{ (k+1)\bar{\zeta}_{k+1}(2-\rho_{n-k}) + BS^2\binom{k+3}{3} \bar{\zeta}_{n-k} \right\}\right]
\end{equation}
\begin{equation}
    \rho_n = 1 - (1-S)^{n+1}.
\end{equation}    
\label{eq:ConvPowerSeries_ODETransformed_prefactor}
\end{subequations}
In writing~(\ref{eq:BesselEuler_Initial}), we use the fact that the Bessel function prefactor does not introduce new singularities that impact convergence, since the actual function that is Euler-transformed and then subsequently expanded in a power series is $[\bar{h}(\Tilde{r})] [K_0(\sqrt{B}(\Tilde{r}+1))^{-1}]$. It is well known~\citep{Parnes} that the zeros of $K_0$ occur in the left-half plane, which means any value of $S$ that successfully Euler sums $\Bar{h}$ must also do so for the function $[\bar{h}(\Tilde{r})] [K_0(\sqrt{B}(\Tilde{r}+1))^{-1}]$. 

To utilize~(\ref{eq:ConvPowerSeries_ODETransformed}), the estimate for the radius of convergence, $S$, is needed (recall that Appendix~\ref{sec:ClosestSing} indicates that S does not need to be exact).  Note that the Domb-Sykes plot could be used to establish its value exactly for given parameter values $\theta$ and $B$. However, to simplify implementation, a systematic estimate may be obtained by considering the asymptotic limits of small and large Bond numbers. When the Bond number is small ($B\rightarrow0$), both sides of~(\ref{eq:RearrangedODE}) can be integrated once, and is thus reduced to a $1^\mathrm{st}$-order ordinary differential equation  as
\begin{equation*}
   \displaystyle \bar{h}' = \frac{-\cos\theta}{\sqrt{(\tilde{r}+1)^2 - \cos^2\theta}},
   \label{eq:ODE_NoGravity}
\end{equation*}
which has a singularity at the location where the denominator is zero and the slope is infinite. The singularity lies on the negative $\tilde{r}$ axis, whose precise location depends on $\theta$. The location of infinite slope imposes a radius of convergence on the physical domain from the closest root to the $\tilde{r} = 0$ location, and the radius of convergence imparted is given by
\begin{equation}
    \tilde{r}_{s}|_{B\rightarrow0}= |1-|\cos\theta||,~~B \ll1.
    \label{eq:rocsmall}
\end{equation}

To find~$\tilde{r}_s$ of~(\ref{eq:DivPowerSeries}) as $B\rightarrow \infty$, one can interpret the problem dimensionally as that of a cylinder with a large radius in a fluid with fixed physical properties. In such a limit, the cylinder surface may be viewed as flat over a large portion of the domain. The meniscus of a flat wall satisfies an autonomous ODE and has an analytical solution~\citep{Batchelor}. The convergence limiting singularity corresponds to the location where the slope of the interface solution is infinite even if outside the physical domain. That singularity is imparted because all flat wall interface solution lie along the same master curve that is translated to meet conditions at the wall. This location is given by~\cite{FlatWallSakiadisPaper} and is rewritten below in terms of current notation and non-dimensionalized variables as
\begin{equation}
    \tilde{r}_{s}|_{B\rightarrow\infty}= \frac{\left|\cosh^{-1}\sqrt{2} - \cosh^{-1}\sqrt{\frac{2}{1-\sin\theta}} + \sqrt{2+2\sin\theta} - \sqrt{2}\right|}{\sqrt{B}},~~B\gg 1.
    \label{eq:roclarge} 
\end{equation}

\begin{figure}[H]%
    \centering
    \includegraphics[width=9cm]{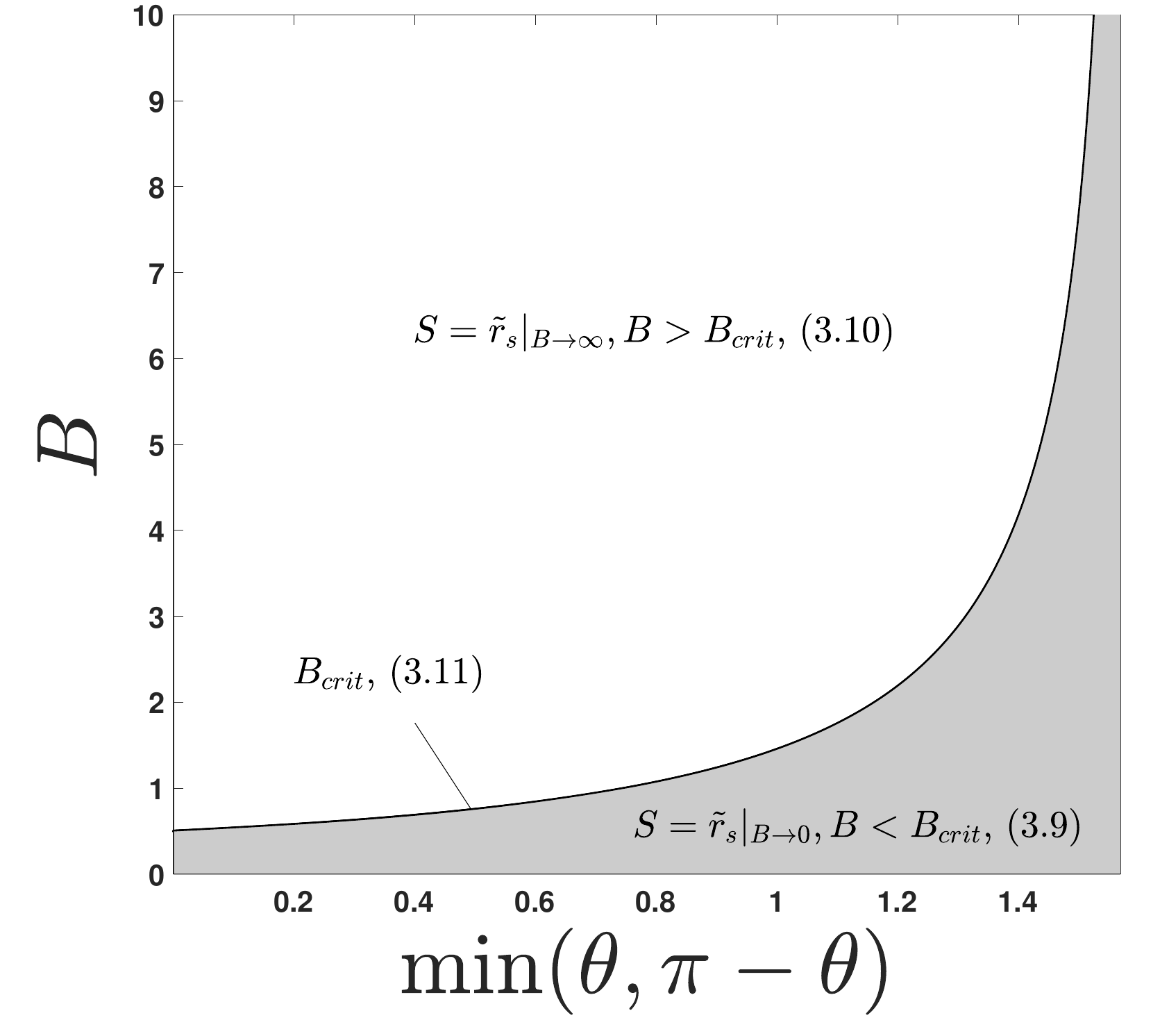}%
    \caption{Determination of $S$ using the relationship given by~(\ref{eq:explicitB}). For any combination of $B$ and $\theta$ values above or below the solid line, we choose~(\ref{eq:roclarge}) or~(\ref{eq:rocsmall}), respectively, to calculate the corresponding radius of convergence, $\tilde{r}_s$, for its use in~(\ref{eq:ConvPowerSeries_ODETransformed}) and~(\ref{eq:ConvPowerSeries_ODETransformed_prefactor}). For any combination of $B$ and $\theta$ values on the solid line, $B_{crit}$ from~(\ref{eq:explicitB}), one may choose to use either of the equations. The plot is not needed for $\theta = \pi / 2$ because the solution is horizontal for all Bond numbers.}%
    \label{fig:Bcrossover}%
\end{figure}

We have surveyed the parameter ranges $\theta\in\left(0, \pi \right)$ and $B\in[0.001,100]$, and find that the ratio of $\tilde{r}_{s}$ found numerically (using a numerical $a_0$ to initiate the recursion in~(\ref{eq:DivPowerSeries})) to the estimated $\tilde{r}_{s}$ found using~(\ref{eq:rocsmall}) and~(\ref{eq:roclarge}) always lies between 0 and 2 when chosen judiciously as follows---this satisfies the constraints for a convergent Euler summation (see Appendix~\ref{sec:ClosestSing}). We have found that the maximum error in the radius of convergence in using either~(\ref{eq:rocsmall}) or~(\ref{eq:roclarge}) is obtained when the radii predictions are equal. Setting~(\ref{eq:rocsmall}) equal to~(\ref{eq:roclarge}) leads to a master curve to solve for a critical value of $B$, denoted by $B_{crit}$, as a function of contact angle $\theta$ as
\begin{equation}
    \displaystyle B_{crit} = \left[ \frac{ \cosh^{-1}\sqrt{2} - \cosh^{-1}\sqrt{\frac{2}{1-\sin\theta}} + \sqrt{2+2\sin\theta} - \sqrt{2}}{1-|\cos\theta|}\right]^{2},
    \label{eq:explicitB}
\end{equation}
which is plotted in figure~\ref{fig:Bcrossover}. In this figure, errors in the radius of convergence (compared with the numerical solution) become smaller as one moves away from the plotted curve.  We can use this fact, then, to determine the estimate for the radius of convergence, $S$, in the Euler transformation~(\ref{eq:EulerTrans}) which is used in the convergent series solutions~(\ref{eq:ConvPowerSeries_ODETransformed}) and~(\ref{eq:ConvPowerSeries_ODETransformed_prefactor}).  Consistent with the derivation,~(\ref{eq:roclarge}) is used to determine $S$ for $B > B_{crit}$, and~(\ref{eq:rocsmall}) is used to predict $S$ for $B < B_{crit}$. For any combination of $B$ and $\theta$ values that fall along the solid line in figure~\ref{fig:Bcrossover},  $\tilde{r}_s$ (i.e., $S$) obtained from~(\ref{eq:rocsmall}) and~(\ref{eq:roclarge}) are identical, but~(\ref{eq:rocsmall}) is preferred for simplicity. In the results and figures that follow, we use~(\ref{eq:rocsmall}) and~(\ref{eq:roclarge}) along with figure~\ref{fig:Bcrossover} to obtain $S$ analytically.

\section{Convergent Resummation Results} \label{sec:ConvergentResults}
\begin{figure}[H]%
    \centering
    \subfloat{\hspace{-0.25in} \label{fig:Error_vs_r_FixedTheta_ODETransformed} \includegraphics[width=8cm]{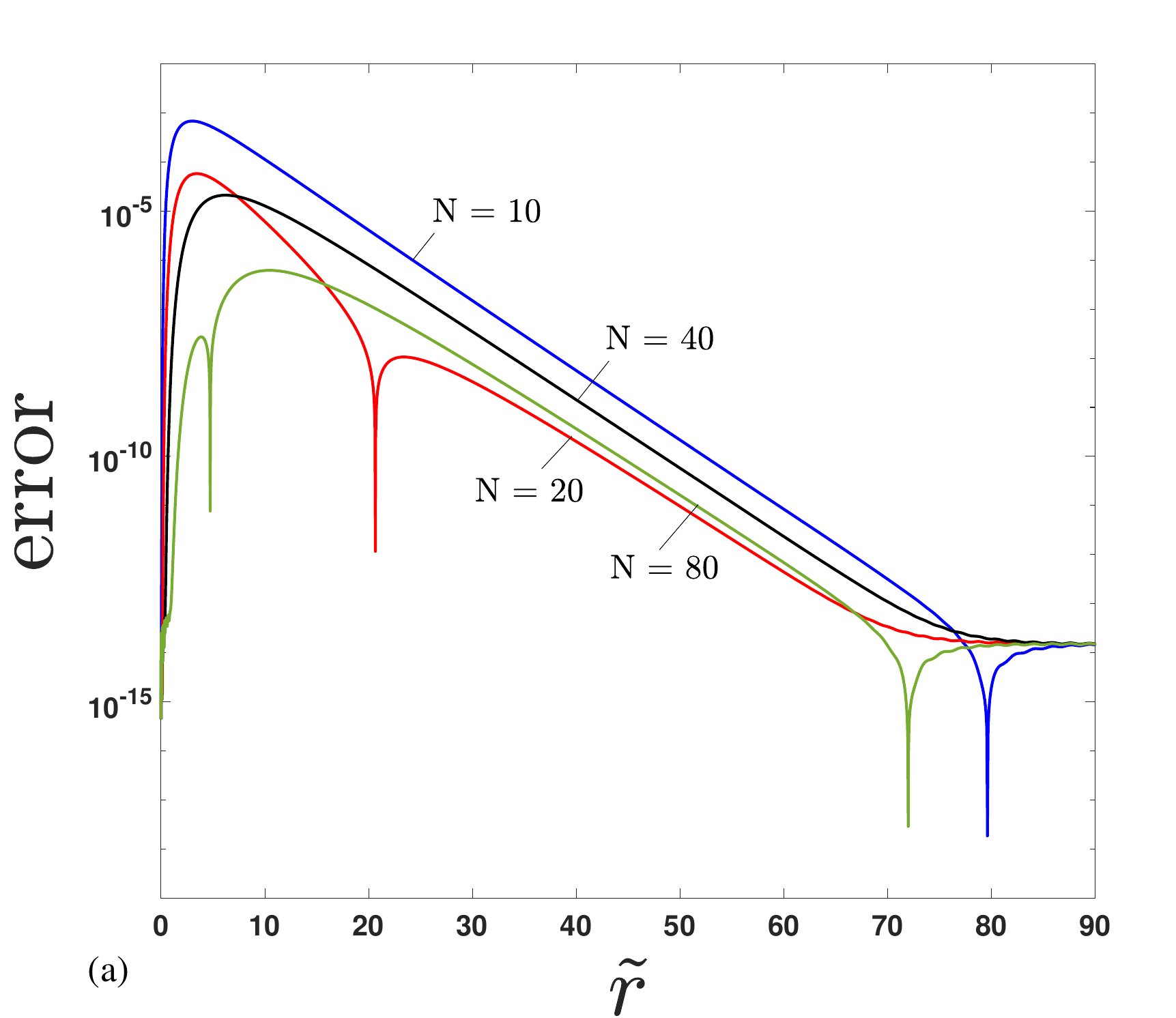}}
    \subfloat{\hspace{-0.15in} \label{fig:Error_vs_r_FixedTheta_NMa0}
    \includegraphics[width=8cm]{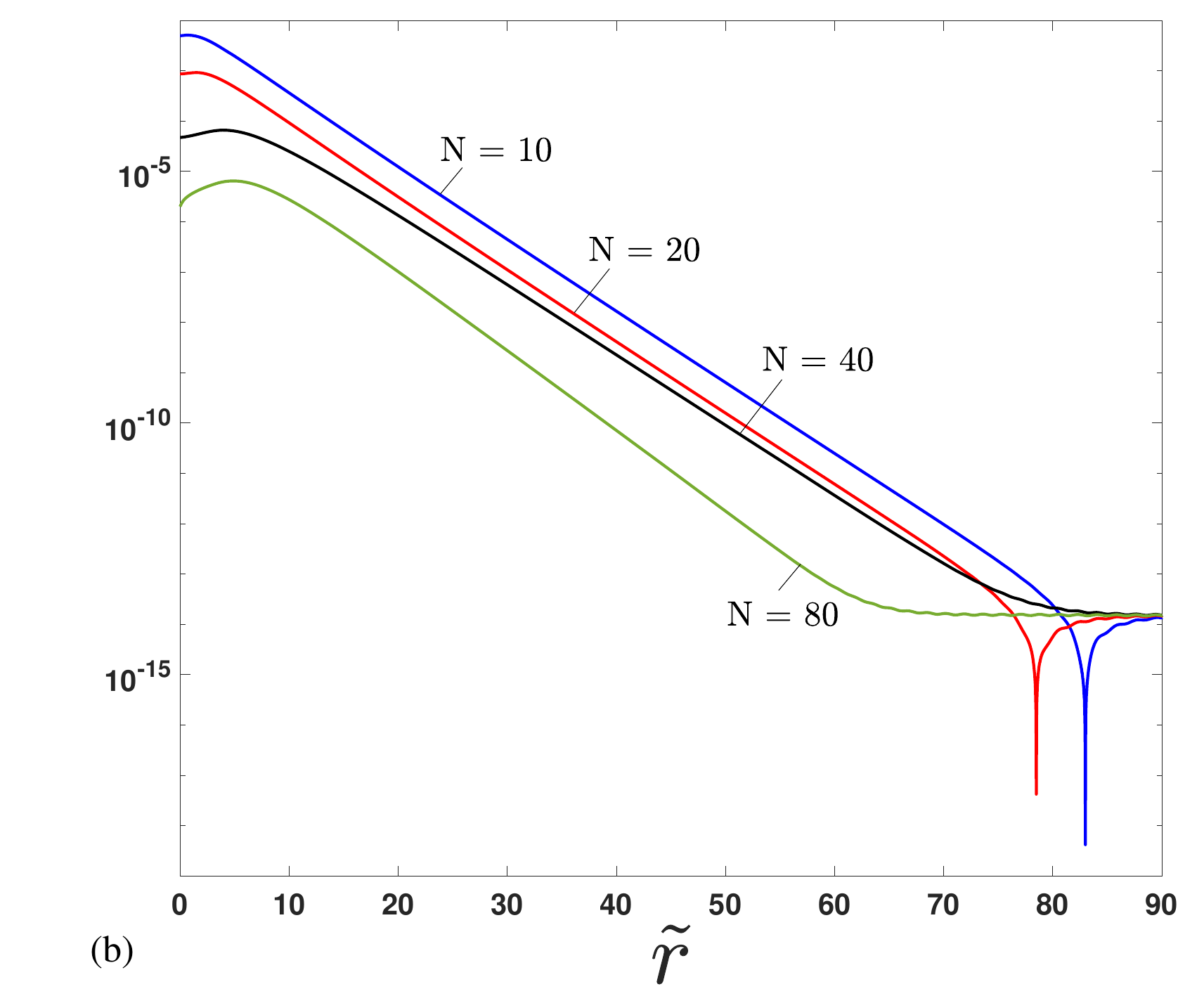}}
    \caption{Absolute error (the absolute value of the difference) between $N$-term truncations of the convergent series solution~(\ref{eq:ConvPowerSeries_ODETransformed_prefactor}) and the numerical solution with a domain length $\mathcal{L}=240$ for $B = 0.1$ and $\theta = \pi/4$ plotted versus $\tilde{r}$ using (a) $a_0 = \bar{h}(\tilde{r}=0)$ generated by the numerical solution, and (b) the predicted value of $a_0$ using the power series itself following the algorithm in section~\ref{sec:HeightPrediction_ODETransformed}.}
\end{figure}

\begin{figure}[H]%
    \centering
    \subfloat{\hspace{-0.25in} \label{fig:InfNorm_vs_N_FixedB_Numa0_ODETransformed} \includegraphics[width=8.1cm]{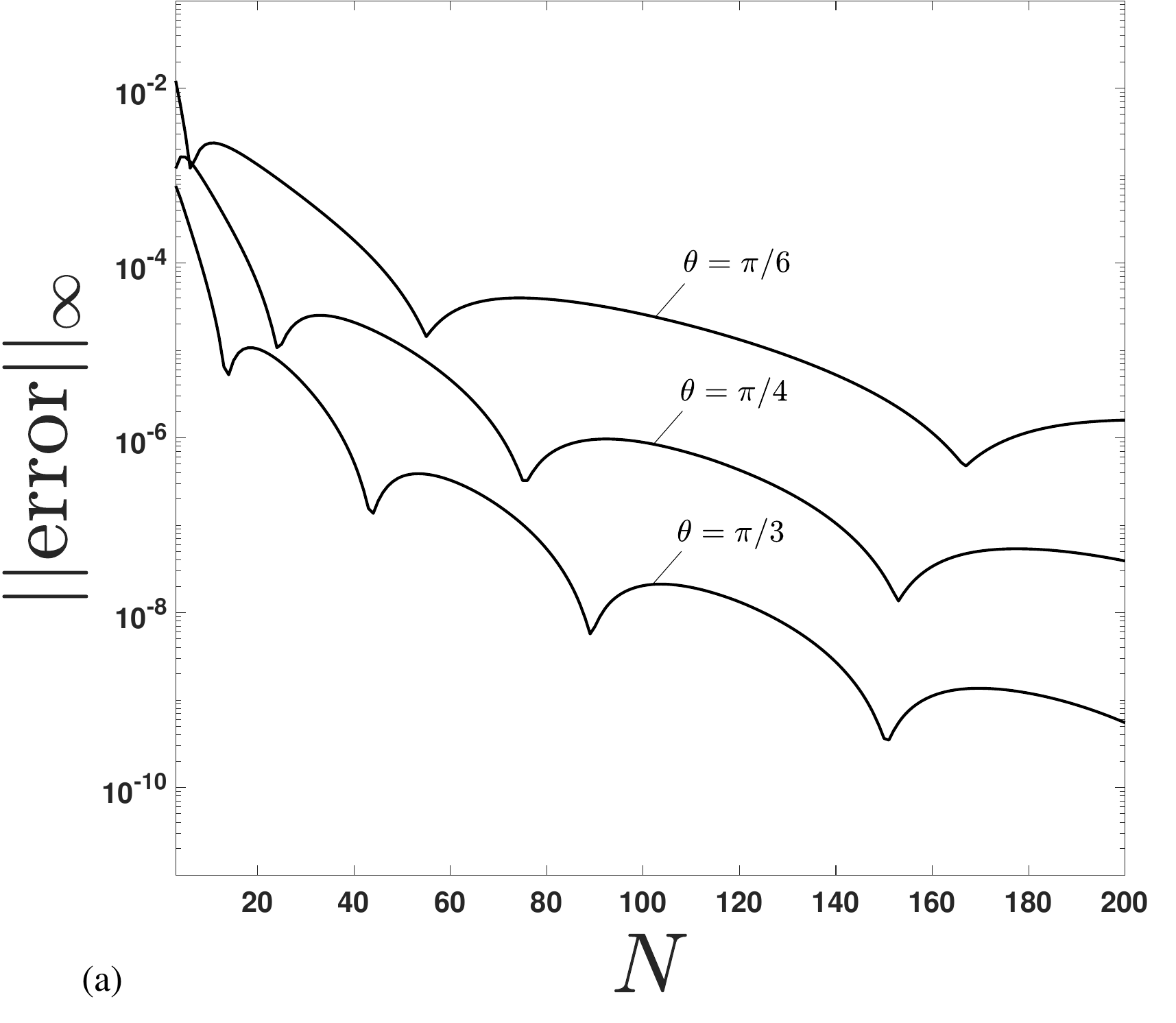}}
    \subfloat{\hspace{-0.11in} \label{fig:Error_vs_N_FixedB_Log_OptionC}
    \includegraphics[width=8.1cm]{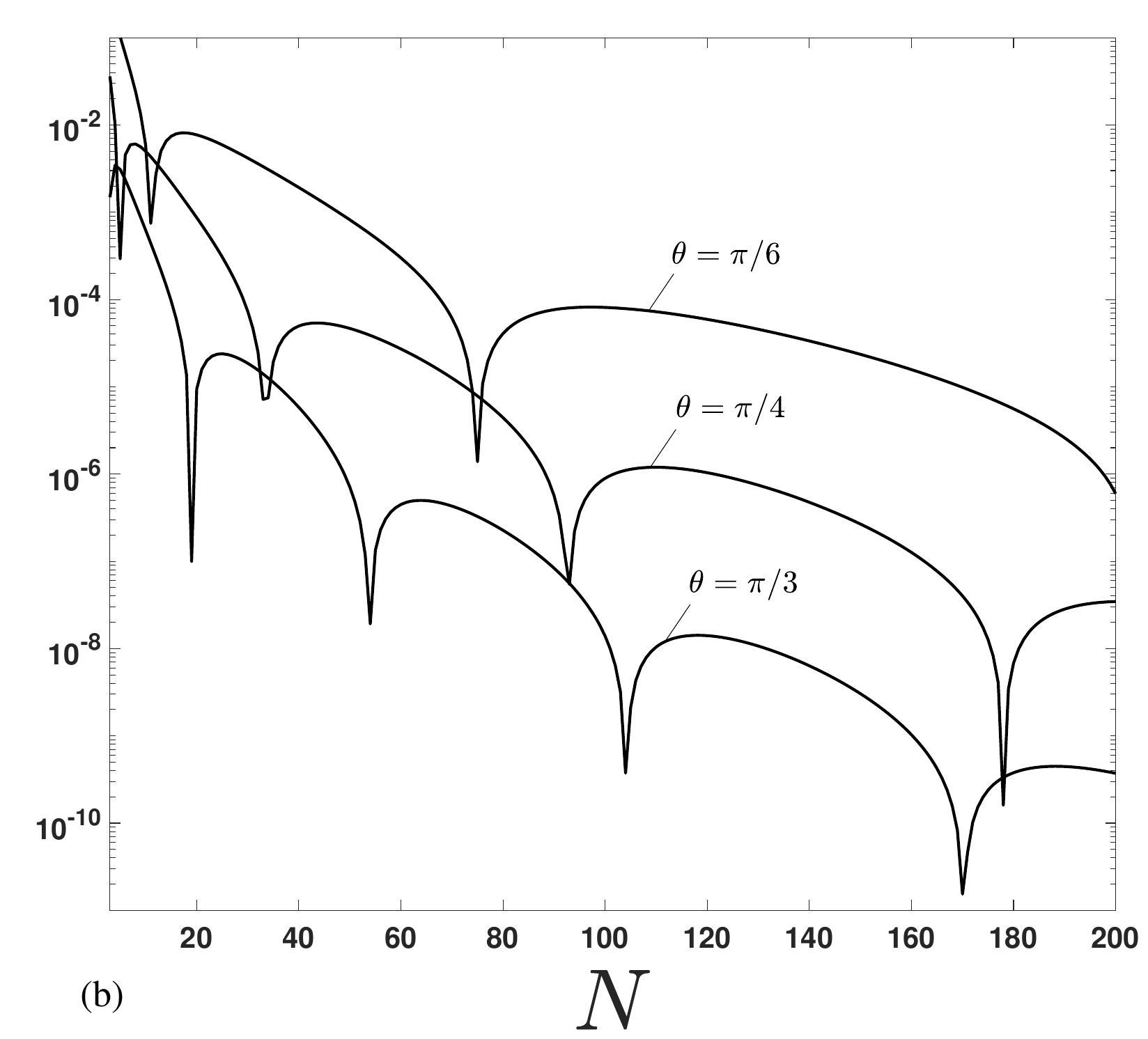}}
    \caption{Maximum absolute error (maximum of the absolute value
    of the difference) taken over $\tilde{r} \in [0,\mathcal{L}]$ between $N$-term truncations (shown by  $\bullet$'s) of the convergent resummation~(\ref{eq:ConvPowerSeries_ODETransformed_prefactor}) and the numerical solution for $B = 0.1$ and $\displaystyle \theta = \pi/6$ (with $\mathcal{L}=240$), $ \pi/4$ (with $\mathcal{L}=240$), and $\pi/3$ (with $\mathcal{L}=240$), plotted versus $N$, using (a) $a_0 = \bar{h}(\tilde{r}=0)$ generated from the numerical solution, and (b) the predicted value of $a_0$ using the power series itself following the algorithm in section~\ref{sec:HeightPrediction_ODETransformed}.}
\end{figure}

\begin{figure}[H]%
    \centering
    \subfloat{\hspace{-0.2in} \label{fig:InfNorm_vs_N_FixedTheta_Numa0_ODETransformed} \includegraphics[width=8cm]{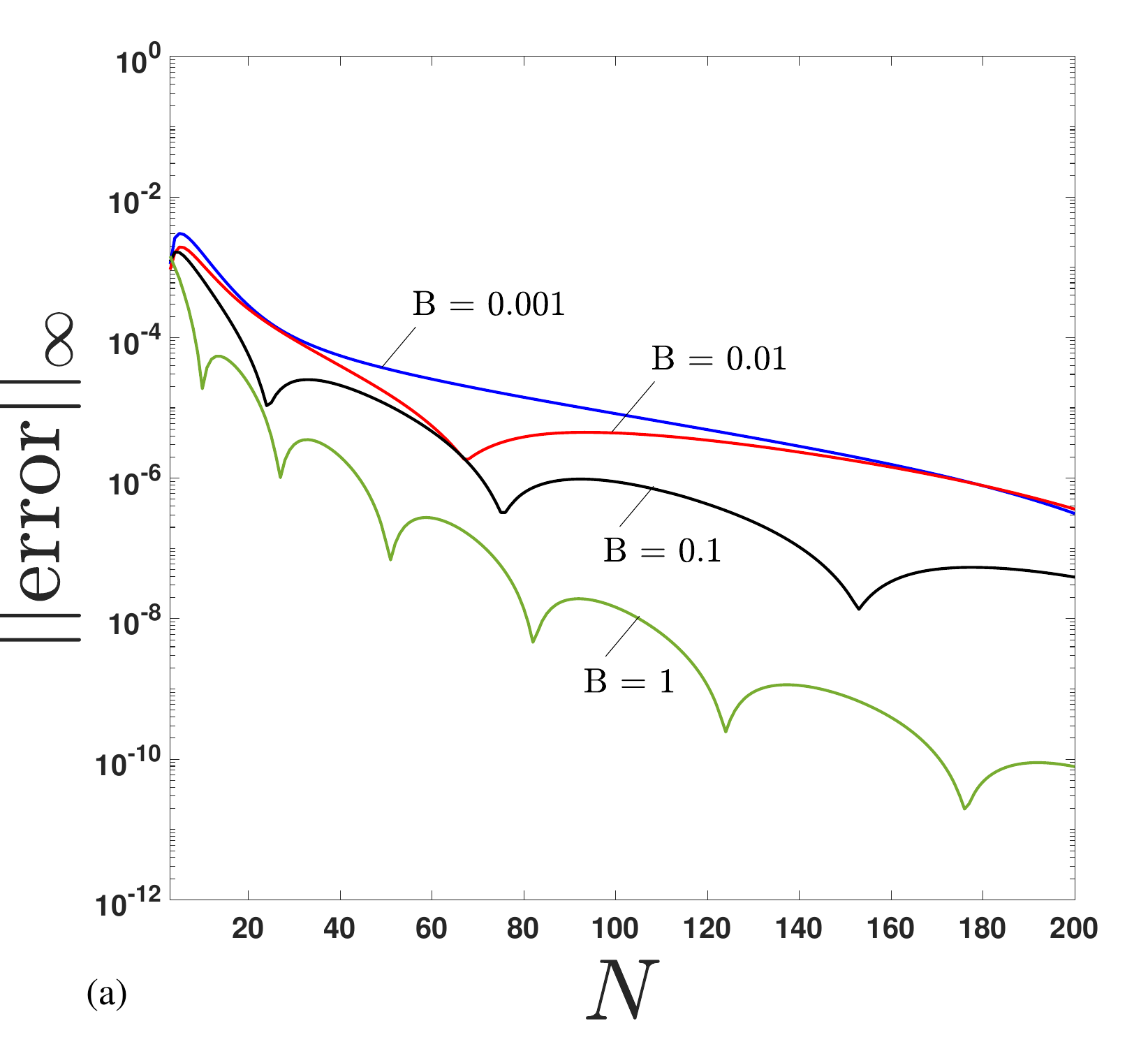}}
    \subfloat{\hspace{-0.15in} \label{fig:InfNorm_vs_N_FixedTheta}
    \includegraphics[width=8cm]{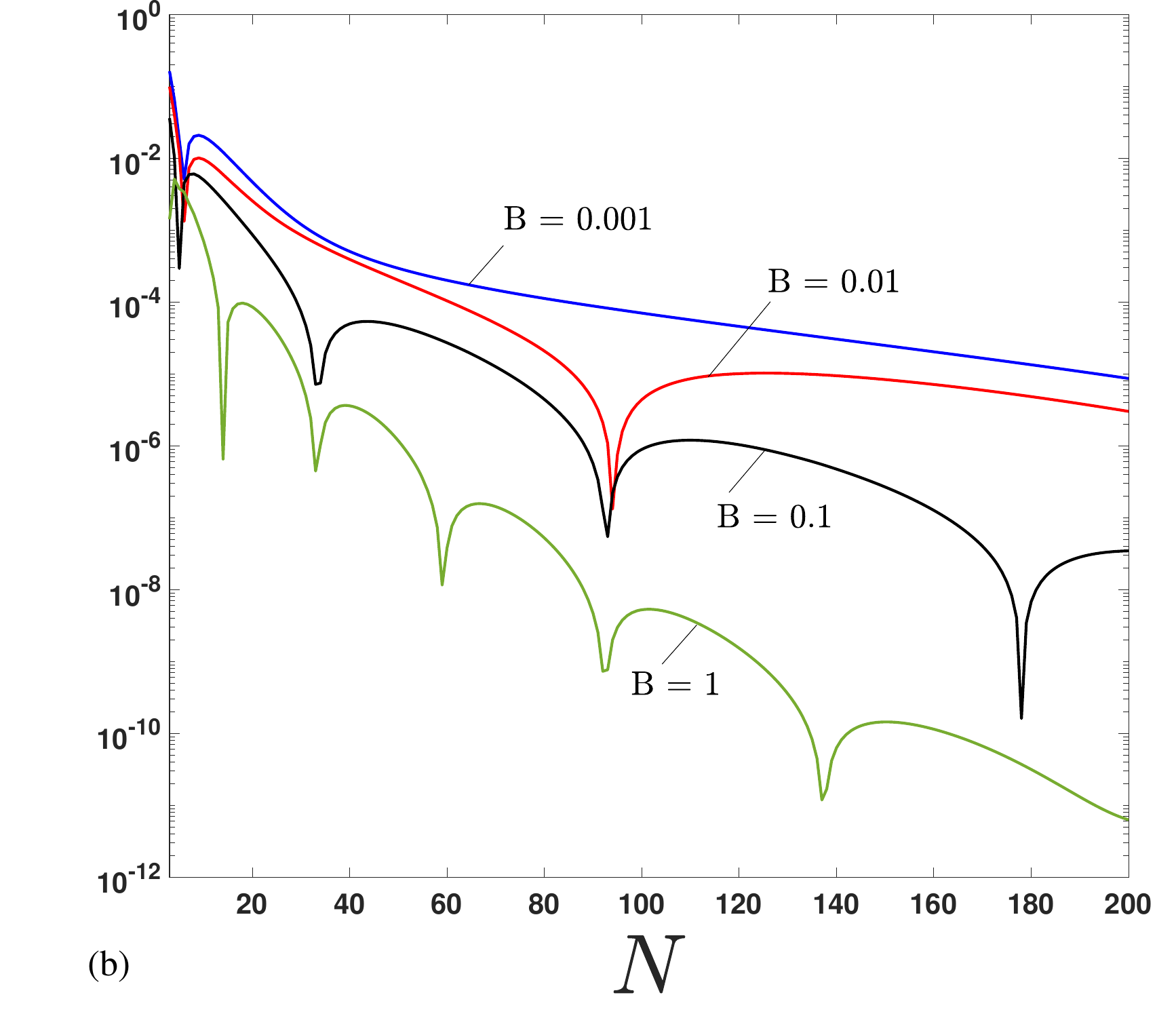}}
    \caption{Maximum absolute error taken $\tilde{r} \in [0,\mathcal{L}]$ between $N$-term truncations (shown by $\bullet$'s) of the convergent resummation~(\ref{eq:ConvPowerSeries_ODETransformed_prefactor}) and the numerical solution for $B= 1$ (with $\mathcal{L}=60$), 0.1 (with $\mathcal{L}=60$), 0.01 (with $\mathcal{L}=480$), and 0.001 (with $L=960$), and $\theta = \pi/4$, plotted versus $N$, using (a) $a_0 = \bar{h}(\tilde{r}=0)$ generated by the numerical solution, and (b) the predicted value of $a_0$ using the power series itself following the algorithm in section~\ref{sec:HeightPrediction_ODETransformed}.}
\end{figure}

In this section, we examine the performance of the series~(\ref{eq:ConvPowerSeries_ODETransformed_prefactor})  by direct comparison with numerical results. We begin by noting that the system~(\ref{eq:FirstScaledODESystem}) has a reflection property that the solutions for $\theta < \pi/2$ are identical in shape to those for that $\theta > \pi/2$ provided $\mid\theta-\pi/2\mid$ are the same. The only difference is that the solutions are mirror images, so solutions rising above the horizontal pool as $r \to \infty$ for $\theta<\pi/2$ fall below that same horizontal height. Thus, it suffices to only consider solutions for $\theta<\pi/2$ in what follows. Figure~\ref{fig:Error_vs_r_FixedTheta_ODETransformed} shows the absolute error (the absolute difference)
between $N$-term truncations of the convergent series solution~(\ref{eq:ConvPowerSeries_ODETransformed_prefactor}) and the numerical solution for $B = 0.1$ and $\theta = \pi/4$. The value of $\bar{h}(\tilde{r}=0)$ from the numerical solution is used as the value of $a_0$ in constructing figure~\ref{fig:Error_vs_r_FixedTheta_ODETransformed} (as well as figures~\ref{fig:InfNorm_vs_N_FixedB_Numa0_ODETransformed}  and~\ref{fig:InfNorm_vs_N_FixedTheta_Numa0_ODETransformed} in this section). Thus, the lowest absolute error (the absolute value of the difference between the convergent series solution~(\ref{eq:ConvPowerSeries_ODETransformed_prefactor}) and the numerical solution) in figure~\ref{fig:Error_vs_r_FixedTheta_ODETransformed} is observed at the wall. Note that the indicated errors are continually reduced for all $\Tilde{r}$ as $N$ is increased beyond those values shown in the figures, but the trends indicated are maintained. Thus, desired accuracy can be achieved with sufficiently large $N$. 

Figure~\ref{fig:InfNorm_vs_N_FixedB_Numa0_ODETransformed} shows the maximum absolute error (maximum of the absolute value of the difference over $\tilde{r} \in [0,\mathcal{L}]$) between the power series representation~(\ref{eq:ConvPowerSeries_ODETransformed_prefactor}) and the numerical solution for $B = 0.1$ and various $\theta$ values, while  figure~\ref{fig:InfNorm_vs_N_FixedTheta_Numa0_ODETransformed} shows the same for $\theta = \pi/4$ and various $B$ values. We observe that the error between the numerical and series solution decreases as contact angles approach $\pi/2$ for the same number of terms. Similarly, for the same number of terms the power series solution~(\ref{eq:ConvPowerSeries_ODETransformed_prefactor}) 
becomes more accurate as the Bond number increases. Again, increased accuracy may be obtained with increasing $N$ beyond those shown in the figures. Moreover, we note that the series takes more terms to converge as $\theta \to 0$, and less terms as $\theta \to \pi/2$, for all Bond numbers.

\section{Prediction of the Meniscus Height at the Cylinder Wall} \label{sec:HeightPrediction_ODETransformed}
Up to this point in the paper, all discussed power series results have utilized the numerically predicted value of the height at the wall as an input. We now provide the methodology to determine the height of the wall, $a_0 = \bar{h}(\Tilde{r}=0)$ using the power series itself, which precludes the need for numerical inputs altogether. The recurrence for the coefficients provides data about the structure of the solution as a function of the chosen value of $a_0$. While there are many conditions we impose on the coefficients of~(\ref{eq:ConvPowerSeries_ODETransformed_prefactor}) to yield an equation for $a_0$, the simplest is to set the last $\bar{A}_n$ coefficient to zero. This approach has been used previously in~\cite{Barlow:2017}. The rationale for this choice is based on the fact that, in a convergent series, the last coefficient, $\bar{A}_N$ must approach zero as $N\to\infty$; thus the assumption that $\bar{A}_N=0$ is self consistent in this limit.  From a purely mathematical perspective, setting $\bar{A}_N=0$ provides one equation in one unknown $a_0$ that enables solution. We can compute the $n$-th coefficient $\bar{A}_n(a_0)$ as a function of the chosen value of $a_0$. Since the underlying recurrence is analytic, we can differentiate it to compute derivatives with respect to $a_0$ as well. As such, we can use Newton's iteration to obtain the desired result by choosing $a_0$ to enforce the condition $\bar{A}_N = 0$. 

The Newton's iteration takes the form 
\begin{subequations}
\begin{equation}
    a_{0,j+1} = a_{0,j} - \frac{\Bar{A}_{N}(a_{0,j})}{\Bar{A}'_{N}(a_{0,j})},
    \label{eq:NewtonsMethod}
\end{equation}
where $\Bar{A}'_{N}(a_{0,j})$ is the derivative of $\bar{A}_N$ with respect to $a_0$. To utilize~(\ref{eq:NewtonsMethod}), we differentiate the coefficients in~(\ref{eq:ConvPowerSeries_ODETransformed_prefactor}) with respect to $a_0$, and obtain
\begin{equation}
   \bar{A}'_n= \sum_{k=0}^n \zeta_{k}  \hat{A}'_{n-k},
\end{equation}    
\begin{equation}
    \hat{A}'_{n+2} = \frac{S^2}{(n+1)(n+2)}\sum_{k=0}^n \binom{k+3}{3} w'_{n-k}, 
\end{equation}
\begin{equation}
    w'_n = f'_n + \sum_{k=0}^n r_{n-k}\left(q'_k - t'_k - u'_k\right)~~,~~f'_n = \frac{2}{S^2}\sum_{k=0}^n (-1)^k \binom{3}{k} (n-k+1) \hat{A}'_{n-k+1},
\end{equation}    
\begin{equation}
     q'_n = B \sum_{k=0}^n \left\{\ell'_k p_{n-k} + \ell_k p'_{n-k}\right\}~~,~~     \ell'_n = \sum_{k=0}^n \bar{\ell}_k \hat{A}'_{n-k},
\end{equation}
 \begin{equation}
     u'_n = \frac{1}{S}\sum_{k=0}^n (-1)^k\binom{2}{k}(n-k+1) \hat{A}'_{n-k+1},
 \end{equation}
\begin{equation}
    d'_n = \sum_{k=0}^n  \left\{u'_k u_{n-k} + u_k u'_{n-k}\right\}~~,~~ t'_n= \sum_{k=0}^n \left\{d'_ku_{n-k} + d_k u'_{n-k}\right\}~~,~~\Bar{d}'_n = d'_n,
\end{equation}
\begin{equation}
    p'_n = \frac{1}{n\bar{d}_0} \sum_{k=1}^n\left\{\left(\frac{5}{2}k-n\right)\left( \bar{d}'_{k} p_{n-k} + \bar{d}_k p'_{n-k}\right) \right\} - \frac{1}{n \bar{d}_0^2} \bar{d}'_0 \sum_{k=1}^n \left(\frac{5}{2}k-n\right) \bar{d}_k p_{n-k}, 
\end{equation}
\begin{equation}
    p'_0 = \frac{3}{2}\bar{d}_0^{1/2} \bar{d}'_0,
\end{equation}
 with
\begin{equation}
    \hat{A}'_0= 1,~\textrm{and}~ \hat{A}'_1 = 0. 
\end{equation}
\label{eq:NewtonMethodAlgorithm}
\end{subequations}
Using an arbitrary initial guess,~(\ref{eq:NewtonMethodAlgorithm}) is used to predict $a_0$, which approaches the correct value as $N \to \infty$, as evidenced by the error plots in figures~\ref{fig:Error_vs_r_FixedTheta_NMa0},~\ref{fig:Error_vs_N_FixedB_Log_OptionC}, and~\ref{fig:InfNorm_vs_N_FixedTheta}. 

Apart from comparing directly with numerical results, our predictions of the height of the interface at the wall ($\Bar{h}=a_0$) and resulting interface shape agree well with an analytical expression derived for small interfacial slopes (see Appendix~\ref{sec:SmallSlope}) over the whole radial domain.  This occurs when $\theta$ is near $\pi / 2$. Under such circumstances, the constant $D$ in~(\ref{eq:prefactor}) can be determined based on constraints at the wall.  According to~(\ref{eq:ConvPowerSeries_ODETransformed_prefactor}), this means that $\bar{A}_0=D$ for small slopes.  If one utilizes the algorithm~(\ref{eq:NewtonMethodAlgorithm}) to determine the constant $A_0$ by setting $A_1=0$, we obtain the same value of the constant $D$ obtained the small slope approximation. This provides additional insight into how the series~(\ref{eq:ConvPowerSeries_ODETransformed_prefactor}) with the modified Bessel prefactor accelerates convergence.  For small slopes at the wall, one term in the expansion provides an excellent prediction of the interface shape, while higher order terms in $\eta$ contribute only incremental improvement. Far from the wall, the same comment applies---as the small slope approximation~(\ref{eq:prefactor}) is always valid, the Bessel function prefactor in~(\ref{eq:ConvPowerSeries_ODETransformed_prefactor}) assures that only few terms in the expansion in $\eta$ (or equivalently $\tilde{r}$) are needed to capture the solution behavior.
\section{Performance of Power Series and Matched Asymptotic solutions\label{sec:AsymptoticVsSeries}} 
\begin{figure}[htbp]%
    \centering
    \includegraphics[width=9cm]{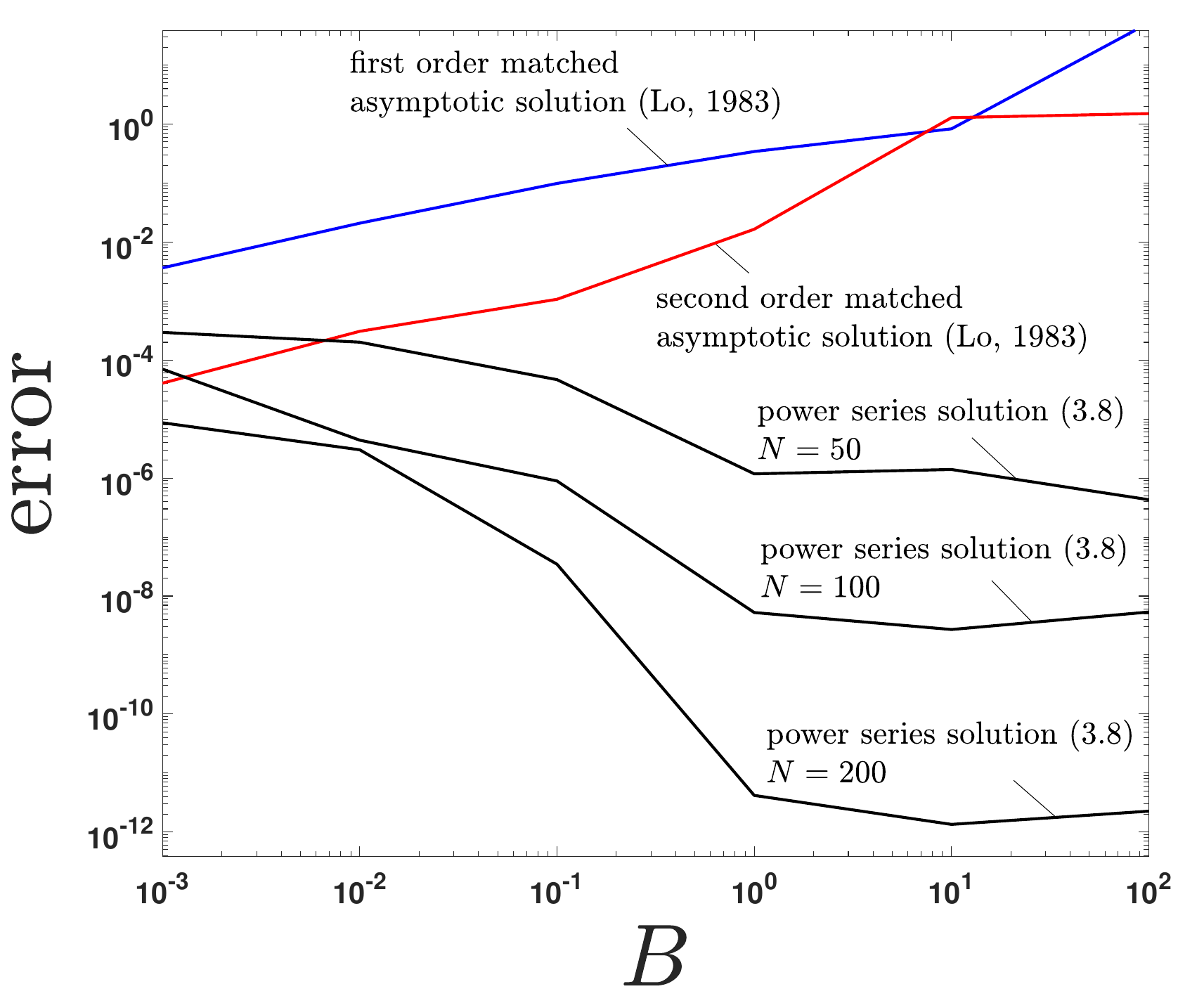}%
    \caption{Absolute error between predictions of the height of the meniscus at the wall of the cylinder ($\tilde{r}=0$) using the power series~(\ref{eq:ConvPowerSeries_ODETransformed_prefactor}) and the numerical solution, compared with error from the first~(\ref{eq:FirstH0}) and second~(\ref{eq:SecondH0}) order matched asymptotic solutions of ~\citep{Lo} for $\theta = \pi/4$.}%
    \label{fig:WallError_vs_B_ODETransformed}%
\end{figure}

In this section, we compare predictions of the height of the interface at the wall, $a_0$ from the power series solution~(\ref{eq:ConvPowerSeries_ODETransformed_prefactor}) using the algorithm of section~\ref{sec:HeightPrediction_ODETransformed},  with those of the matched asymptotic solution by~\cite{Lo} in the limit $B\rightarrow 0$. Figure~\ref{fig:WallError_vs_B_ODETransformed} compares the solutions generated by the $N$-term truncation of the convergent power series solution~(\ref{eq:ConvPowerSeries_ODETransformed_prefactor}), the first order matched asymptotic solution of~\cite{Lo}, and the second order matched asymptotic solution of~\cite{Lo} (see~(\ref{eq:FirstH0}) and~(\ref{eq:SecondH0}) in Appendix~\ref{sec:MatchedAsympSol}, respectively), with the numerical solution for $B=$ 100, 10, 1, 0.1, 0.01, and 0.001 and $\theta=\pi/4$, at the wall of the cylinder ($\tilde{r} = 0$). 

As shown in figure~\ref{fig:WallError_vs_B_ODETransformed}, the accuracy of the asymptotic solution improves as $B$ approaches zero, as expected from asymptotic theory. That said, the series solution~(\ref{eq:ConvPowerSeries_ODETransformed_prefactor}) is effective over a broad range of Bond numbers, and in fact performs with high accuracy even in the range where asymptotic analysis is justified; that is, provided a sufficient number of terms in the series are included. 

In the case of small Bond numbers where Lo's solution has sufficient accuracy, one could choose to use the value of $a_0$, the height of the liquid meniscus at the wall, from the method of matched asymptotic solutions~\citep{Lo}, shown in~(\ref{eq:FirstH0}) and~(\ref{eq:SecondH0}), and use the convergent series solution~(\ref{eq:ConvPowerSeries_ODETransformed_prefactor}) to generate the full shape of the interface. The accuracy of the matched asymptotic solution might be improved by considering a third order match, but as is evident from the work of~\cite{Lo}, the generation of higher order matches is arduous. Additionally, asymptotic series are subject to optimal truncation errors, so there is a limit to the improvement that may be achieved by adding such corrections. By contrast, the power series approach yields a relatively simple recursion that can be used for solutions of desired accuracy by simply including additional terms in the series. 
\section{Conclusion} \label{sec:Conclusion} 
In this work, the asymptotically motivated Euler transformation was utilized to transform a divergent power series solution to a convergent one, where the influence of singularities were mapped outside the physical domain. Furthermore, we have obtained---for the first time---an exact solution for the shape of the meniscus formed outside a partially submerged vertical cylinder in an infinite horizontal static pool. The efficacy of the power series solution has been systematically tested against numerical and previously-derived matched asymptotic solutions. Moreover, the complexity of the matched asymptotic solution precludes additional higher-order corrections from being generated practically. The power series solution, generated recursively, is attractive for its ease of use 
for all values of $B$, $\theta$, and over the entire physical domain.
\appendix
\numberwithin{equation}{section}
\section{Matched Asymptotic Solution} \label{sec:MatchedAsympSol}
In this section, we re-visit the first and the second order matched asymptotic solutions implemented by~\cite{Lo} and convert their notation to ours. In doing so, we correct two typos in Eqns 3.18 and 3.21 of Lo's paper~\citep{Lo}, shown respectively with~\underline{underlines} in equations~(\ref{eq:FirstTypo}) and~(\ref{eq:SecondTypo}) below. Here we focus on Lo's predictions of the height of the meniscus along the cylinder to benchmark against the power series solution. Lo uses the superscripts (1) and (2) to represent the first and second order matched asymptotic approximations, respectively, and denotes the well-known Euler's constant as $\gamma = 0.5772$ (rounded to four decimal places). Lo defines the parameter $\epsilon = \frac{R}{l_c}$, where $R$ is the cylinder radius and $l_c = \sqrt{\frac{\sigma}{\rho g}} $ is the capillary length. In the notation used in our paper, this means that the Bond number and $\epsilon$ are related as $B = \epsilon^{2}$.

The lowest (first) order matched asymptotic solution for the height of the fluid along the wall, used in figure~\ref{fig:WallError_vs_B_ODETransformed}, is
\begin{subequations}
\begin{equation}
    \bar{h}^{(1)} (\tilde{r}=0)= \sin\phi\left\{ \ln\frac{4}{\epsilon(1+\cos\phi)}-\gamma\right\},
\end{equation}
\begin{equation}
    \phi=\frac{\pi}{2}-\theta.
\end{equation}
    \label{eq:FirstH0}
\end{subequations}
The second order matched solution (next order match) for the height of the interface along the cylindrical wall indicated in figure~\ref{fig:WallError_vs_B_ODETransformed} is
\begin{subequations}
\label{eq:SecondH0}
\begin{equation}
    \bar{h}^{(2)} (\tilde{r}=0)=\underbrace{z_{1}\ln\epsilon + z_{2}}_{\bar{h}^{(1)}~\textrm{in} \ (\ref{eq:FirstH0})} + z_{3}\epsilon^{2}\ln^{2}\epsilon + z_{4}\epsilon^{2}\ln\epsilon + z_{5}\epsilon^{2}.
    \label{eq:SecondH0a}
\end{equation}
The constants used in~(\ref{eq:SecondH0a}) are defined as
\begin{equation}
    z_{1} = -\sin\phi~~,~~z_{2} = \sin\phi\left[\ln4 - \ln(1+\cos\phi) - \gamma\right]~~,~~z_{3} = \frac{1}{2}\sin\phi (1-\sin^{2}\phi),
\end{equation}
\begin{equation}
    z_{4} = \sin\phi \left\{\cos^{2}\phi\left[\gamma+\frac{1}{4}+\ln(\frac{1}{4} (1+\cos\phi)) + \right]  + \frac{1}{4}-\cos\phi\right\},
\end{equation}
\begin{align}
\nonumber
    \cr z_{5} =&~\frac{1}{4}\sin\phi \cos\phi(\ln4 - \gamma)-\frac{D}{\cos\phi}+\frac{1}{4}\sin\phi \\
    \nonumber
    & +D(1-\ln4 \textcolor{red}{\underline{+}}\gamma)+ \frac{1}{4}\sin^{3}\phi\left(\frac{1}{2}-\gamma^{3} -\ln^{2}2+\gamma\ln4-\ln2 \right)  \\
    \nonumber
    &  + \left[\ln(1+\cos\phi) \right]\left[\frac{1}{4}\sin^{3}\phi\left(\ln4 - \gamma+\frac{1}{\cos\phi}\right)+D-\frac{1}{4}\sin\phi\cos\phi\right]  \\
    &-\frac{1}{4}(\sin^{3}\phi)\ln^{2}(1+\cos\phi),
    \label{eq:FirstTypo}
\end{align}
where
\begin{equation}
    D=\frac{1}{4}\mathcal{C}(2-\mathcal{C}^{2})\ln[1+\sqrt{1-\mathcal{C}^2}] - (1-\mathcal{C}^2)\frac{1}{2}\mathcal{C}(\ln4-\gamma)-\frac{1}{4}\mathcal{C}\sqrt{ 1 - \textcolor{red}{\underline{\mathcal{C}^2}}},
    \label{eq:SecondTypo}
\end{equation}
and 
\begin{equation}
    \mathcal{C}=\sin\phi.
\end{equation}
\end{subequations}
\section{Useful formulae for manipulating series \label{sec:series}}
\subsection{Raising a Series to a Power\label{sec:JCP}}
The following relation is JCP Miller's formula for raising a series to a power~\citep{Henrici}:
 \begin{subequations}
\begin{equation}
    \left(\sum_{n=0}^{\infty}a_{n}x^{n}\right)^\gamma = \sum_{n=0}^{\infty}b_{n}x^{n}, 
\end{equation}
where
\begin{equation}
    b_{n>0} = \frac{1}{n~a_{0}}\sum_{j=1}^{n}(j\gamma-n+j) a_j b_{n-j},~b_0 = \left(a_0\right)^\gamma. 
\end{equation}
\label{eq:JCP}
\end{subequations}
\subsection{Product of Two Series \label{sec:Cauchy}}
The following relation is the well-known Cauchy product of two series~\citep{Churchill}:
\begin{equation}
    \sum_{n=0}^{\infty}a_{n}x^{n}\sum_{n=0}^{\infty}b_{n}x^{n} = \sum_{n=0}^{\infty}\left(\sum_{j=0}^{n}a_{j}b_{n-j}\right)x^{n},~a_0 \neq 0.
    \label{eq:Cauchy}
\end{equation}
\subsection{Generalized Product Rule \label{sec:productrule}}
The generalized product (Leibniz's) rule applied to two functions is given as~\citep{GR}
\begin{equation}
\displaystyle\frac{d^n\left(uv\right)}{dy^n}=\sum_{k=0}^n{{n}\choose{k}}~\frac{d^{n-k}u}{dy^{n-k}}~\frac{d^kv}{dy^k}
\label{eq:productrule}
\end{equation}
and is important in determining the coefficients of an Eulerized series.

\subsection{Chain Rule \label{sec:chainrule}}
The chain rule applied to a composite function $F(y(\eta))=f(\eta)$ is given, for first and second derivatives, as
\begin{subequations}
\begin{equation}
\frac{df}{d\eta}=\frac{dF}{dy}\frac{dy}{d\eta},
\label{eq:chainrule1}
\end{equation}
\begin{equation}
\frac{d^2f}{d\eta^2}=\frac{d^2F}{dy^2}\left(\frac{dy}{d\eta}\right)^2+\frac{dF}{dy}\frac{d^2y}{d\eta^2}.
\label{eq:chainrule2}
\end{equation}
\label{eq:chainrule}
\end{subequations}
The inversion of~(\ref{eq:chainrule}) is given by 
\begin{subequations}
\begin{equation}
\frac{dF}{dy}=\frac{df}{d\eta}\left(\frac{dy}{d\eta}\right)^{-1},
\label{eq:chainrule3}
\end{equation}
\begin{equation}
\frac{d^2F}{dy^2}=\frac{d^2f}{d\eta^2}\left(\frac{dy}{d\eta}\right)^{-2}-~\frac{df}{d\eta}\frac{d^2y}{d\eta^2}\left(\frac{dy}{d\eta}\right)^{-3},
\label{eq:chainrule4}
\end{equation}
\label{eq:inversion}
\end{subequations}
which will be useful in manipulating the Bessel series in terms of the Euler transformed variable in appendix~\ref{sec:Appendix_EuelerizedBessel}.

\subsection{Euler Transformation \label{sec:Euler}}
Given a power series
\begin{equation}
f=\sum_{n=0}^\infty a_n x^n,
\label{eq:Euler:divSeries}
\end{equation}
the Euler Sum is defined as follows
\begin{subequations}
\begin{equation}
\displaystyle f=\sum_{n=0}^\infty b_n \left(\frac{x}{x+S}\right)^n,
\label{eq:Euler:EulerSum}
\end{equation}
where
\begin{equation}
b_{n>0}=\sum_{m=1}^n{{n-1}\choose{m-1}}a_mS^m,~b_0=a_0.
\label{eq:Euler:EulerCoeffs}
\end{equation}
\label{eq:Euler2}
\end{subequations}
Note that~(\ref{eq:Euler:EulerSum}) is a power series in terms of the expansion variable
\begin{equation}
y\equiv x/(x+S),
\label{eq:Euler:y}
\end{equation}
such that the physical domain $x\in[0,\infty)$ maps to the transformed physical domain $y\in[0,1]$.  The derivation of~(\ref{eq:Euler:EulerCoeffs}) is provided below.

First, equating the right-hand-sides of~(\ref{eq:Euler:divSeries}) and~(\ref{eq:Euler:EulerSum}) and using definition~(\ref{eq:Euler:y}) to write in terms of $y$, we have
\begin{equation}
\sum_{m=0}^\infty a_m\left(\frac{yS}{1-y}\right)^m=\sum_{m=0}^\infty b_m y^m,
\label{eq:Euler:equating}
\end{equation}
where a deliberate switch has been made to using the index $m$ instead of $n$ so that we may extract $b_n$ at a specific value within the above (i.e., at $m=n$) as 
\begin{equation}
b_n=\frac{1}{n!}\left\{\frac{d^n}{dy^n}\left[\sum_{m=0}^\infty a_m\left(\frac{yS}{1-y}\right)^m\right]\right\}_{y=0}.
\label{eq:Euler:bn}
\end{equation}
Note that the above expression comes from applying the definition of a coefficient of a Taylor expansion about $y=0$, which is precisely what the right-hand side of~(\ref{eq:Euler:equating}) is. Our task remains to show that the infinite series given by~(\ref{eq:Euler:bn}) may be rewritten as the finite series~(\ref{eq:Euler:EulerCoeffs}). Interchanging the differentiation and summation operators and rewriting the inside as a product,~(\ref{eq:Euler:bn}) becomes 
\begin{equation}
b_n=\frac{1}{n!}\left\{\sum_{m=0}^\infty a_mS^m\frac{d^n}{dy^n}\left[y^m\left(1-y\right)^{-m}\right]\right\}_{y=0}.
\label{eq:Euler:bn2}
\end{equation}
In order to apply the generalized product rule to the $\left[y^m\left(1-y\right)^{-m}\right]$ term above, we first recognize that individual generalized derivatives of the two terms are 
\begin{subequations}
\begin{equation}
    \frac{d^j (y^m)}{dy^j}=\left[\prod_{\ell=0}^{j-1} \left(m-\ell\right)\right] y^{m-j}
\end{equation}
and
\begin{equation}
    \frac{d^j \left[\left(1-y\right)^{-m}\right]}{dy^j}=\left[\prod_{\ell=0}^{j-1} \left(m+\ell\right)\right] \left(1-y\right)^{-m-j}.
\end{equation}
\label{eq:Euler:derivs}
\end{subequations}
Substituting~(\ref{eq:Euler:derivs}) into the generalized product rule~(\ref{eq:productrule}) to evaluate the derivative in~(\ref{eq:Euler:bn2}), we obtain the expression
\begin{align*}
    b_n=& \frac{1}{n!}\left\{\sum_{m=0}^\infty a_mS^m \right.\\ 
    &\left. \sum_{k=0}^n\left[{{n}\choose{k}}  \left[\prod_{\ell=0}^{n-k-1}\!\!\!\!\!\left(m-\ell\right)\right]y^{m-n+k}\left[\prod_{\ell=0}^{k-1}\!\left(m+\ell\right)\right]\left(1-y\right)^{-m-k}  \right] \right\}_{y=0}.
\end{align*}
Finally, we evaluate the above expression at $y=0$ and note that the only nonzero contributions will come from $y^0$ terms in the inner sum, which correspond with an index of $k=n-m$. Since $k$ is a positive index, this tells us that $m$ cannot be greater than $n$, thus placing an upper limit on the outer $m$-sum.  Setting $y=0$, only retaining the $k=n-m$ term in the inner sum, and truncating the outer sum to $n$ terms, leads to 
\begin{align}
    \nonumber
    b_n=&\displaystyle~\frac{1}{n!}\sum_{m=0}^n a_mS^m{{n}\choose{n-m}}~~~~~~ \displaystyle\prod_{\ell=0}^{m-1}\left(m-\ell\right) ~~~~~~\displaystyle\prod_{\ell=0}^{n-m-1}\left(m+\ell\right)\\
    \nonumber
    = &\displaystyle~\frac{1}{n!}\sum_{m=0}^n a_mS^m\frac{n!}{m!\left(n-m\right)!}~~~~~~~~~m!~~~~~~~~~~~~~~~~~ \displaystyle\frac{\left(n-1\right)!}{\left(m-1\right)!}\\
    \nonumber
    = &\displaystyle~\sum_{m=0}^n a_mS^m\frac{\left(n-1\right)!}{\left(m-1\right)!\left(n-m\right)!} 
\\
    \nonumber
    = & \displaystyle~\sum_{m=0}^n{{n-1}\choose{m-1}}a_mS^m,
    \label{eq:EulerSumGamma}
 \end{align}
 which is equivalent to~(\ref{eq:Euler:EulerCoeffs}). 
\subsection{Estimation of the Radius of Convergence in an Euler Transformation}
\label{sec:ClosestSing}
Typically, for nonlinear ODEs, the precise value of the radius of convergence of the power series solution is unknown, but can be estimated (via root test, ratio test, etc.) to within some tolerance. In such cases, we may estimate the true radius of convergence, $x_s$, with an approximation, $S$, as
\begin{equation}
    S\equiv\epsilon x_s,
\end{equation}
where $\epsilon\in(0,2)$.  We are afforded this flexibility from Taylor's theorem, which tells us that, through the mapping from $x$ to $y$ given by~(\ref{eq:Euler:y}), a Taylor expansion about $y=0$ will converge over the full physical domain $y\in[0,1]$ as long as singularities lie outside of the circle $|y|=1$ in the complex $y$ plane.   In other words, assuming the closest singularity is due to $x=-x_s$, the following must hold for convergence of the corresponding Euler series:
\begin{equation}
\left|y\left(x=-x_s\right)\right|>1.
\label{eq:Euler:constraint}
\end{equation}
We know this to be true in the standard Euler series since then the left side of the inequality above becomes $\pm\infty$, from the standard Euler gauge function~(\ref{eq:Euler:y}). However, if we do not know the exact value of $x_s$ and  substitute the Euler transformation 
\begin{equation}
y\equiv x/(x+\epsilon x_s),
\label{eq:Euler:y2}
\end{equation}
into~(\ref{eq:Euler:constraint}), we obtain
\[\left|\frac{1}{1-\epsilon}\right|>1,\]
which tells us that $\epsilon\in(0,2)$ is a valid range to assure convergence of the transformed series. This means that one may use any arbitrarily small value for $S > 0$ in~(\ref{eq:Euler2}), as long as it is less than twice the value of the (unknown) exact $x_s$ (the modulus of the actual closest singularity). That said, as $\epsilon\to1$, the convergence rate of the Eulerized series is expected to improve, since---in this limit---the influence of the singularity is pushed further away from the mapped physical domain.
\section{Interface Solution for Small Slope} \label{sec:SmallSlope}
Equation~(\ref{eq:prefactor}) provides the leading order $\tilde{r}\to \infty $ asymptotic behavior of the interface solution given as:
\begin{equation}
    \overline{h}(\tilde{r})\sim D{{K}_{0}}\left( \sqrt{B}\left( \tilde{r}+1 \right) \right),
    \label{eq:SmallSlope_PreFactor}
\end{equation}
where $D$ is an unknown constant and ${{K}_{0}}\left( z \right)$ is the modified Bessel function of zeroth order. Note that this solution is valid for $d\overline{h}/d\tilde{r}\ll 1$ since $\overline{h}\to 0$ as $\tilde{r}\to \infty $ according to~(\ref{eq:TransformedStaticPoolBC}). In a configuration for which $d\overline{h}/d\tilde{r}\ll 1$ for all $\tilde{r}\in \left(0,\infty  \right)$,~(\ref{eq:SmallSlope_PreFactor}) provides an excellent approximation to the interface solution with the constant, $D$, determined as follows.  

The boundary condition~(\ref{eq:TransformedSlopeBC}) is rewritten here for convenience as:
\begin{equation}
    \frac{d\bar{h}}{d\tilde{r}}=-\cot\theta~\textrm{at} \ \tilde{r}=0.
    \label{eq:SmallSlope_BC_Slope1}
\end{equation}
Noting that:
\begin{equation*}
    \frac{d{{K}_{0}}(z)}{dz}=-{{K}_{1}}(z),
\end{equation*}
where ${{K}_{1}}(z)$ is the modified Bessel function of first order, we differentiate~(\ref{eq:SmallSlope_PreFactor}) to obtain:
\begin{equation}
    \frac{d\bar{h}}{d\tilde{r}}\sim-D\sqrt{B}{{K}_{0}}\left( \sqrt{B}\left( \tilde{r}+1 \right) \right)~\textrm{for}~~d\overline{h}/d\tilde{r}\ll 1.
    \label{eq:SmallSlope_BC_Slope2}
\end{equation}
Provided that $\cot \theta \ll 1$,~(\ref{eq:SmallSlope_BC_Slope2}) is valid for all $\tilde{r}\in \left( 0,\infty  \right)$, and thus the constraint~(\ref{eq:SmallSlope_BC_Slope1}) may be applied  to determine $D$ as:
\begin{equation}
    D=\frac{\cot \theta }{\sqrt{B}{{K}_{1}}\left( \sqrt{B} \right)}.
    \label{eq:SmallSlope_D}
\end{equation}
Finally, we substitute~(\ref{eq:SmallSlope_D}) into~(\ref{eq:SmallSlope_PreFactor}) to yield:
\begin{subequations}
\begin{equation}
        \overline{h}(\tilde{r})\sim\frac{\cot \theta }{\sqrt{B}{{K}_{1}}\left( \sqrt{B} \right)}{{K}_{0}}\left( \sqrt{B}\left( \tilde{r}+1 \right) \right)~\textrm{for}~~\cot \theta \ll 1,~\textrm{where}~~\tilde{r}\in \left( 0,\infty  \right).
        \label{eq:SmallSlope_h}
\end{equation}
In accordance with the restriction that $\cot \theta \ll 1$,~(\ref{eq:SmallSlope_h}) is valid for nearly horizontal interfaces for which values of $\theta$ lie near $\pi / 2$, regardless of the value of the Bond number, $B$. The height of the interface, then at the cylindrical wall $(\tilde{r}=0)$, is thus given as:
\begin{equation}
    \overline{h}(0)\sim \frac{\cot \theta }{\sqrt{B}}\frac{{{K}_{0}}\left( \sqrt{B} \right)}{{{K}_{1}}\left( \sqrt{B} \right)},~\textrm{for}~~\cot \theta \ll 1 .    
\end{equation} 
\label{eq:SmallSlope_h0}
\end{subequations}
The result~(\ref{eq:SmallSlope_h0}) serves as a useful analytical check on numerical and power series predictions for nearly horizontal interfaces. 
\section{Computation of the Euler coefficients in~(\ref{eq:ConvPowerSeries_ODETransformed}) via the transformed differential equation~(\ref{eq:g_transformedODE})}
\label{sec:Appendix_CoeffComputation}
After making the variable transformation~(\ref{eq:EulerTrans}) in our governing ODE~(\ref{eq:RearrangedODE}), we obtain the transformed ODE~(\ref{eq:g_transformedODE}), rewritten here in a slightly different form as \begin{equation}
    g'' = \frac{S^2}{(1-\eta)^4}\left\{\frac{\displaystyle 2(1-\eta)^3 }{\displaystyle S^2}g' + \frac{\displaystyle \left(\frac{\eta S}{1-\eta} + 1 \right)\left(\frac{(1-\eta)^4}{S^2}g'^2+1\right)^{3/2}Bg - \frac{(1-\eta)^6}{S^3}g'^3-\frac{(1-\eta)^2}{S}g'}{\displaystyle \frac{\eta S}{1-\eta}+1 } \right\}. \label{maintransode}
\end{equation}

Our goal is to obtain a power series solution of~(\ref{maintransode}) in the form~(\ref{eq:gseries}), rewritten here for convenience as
\begin{equation}
    g(\eta) = \sum_{n=0}^\infty \hat{A}_{n}\eta^n.
    \label{eq:gseriesD}
\end{equation}
To do so, we note that, by direct differentiation, the first $n$ terms of the Taylor series of $g'$ depend only on the first $n+1$ terms of $g$. By Cauchy's Product rule and JCP Miller's Formula, the same is also true of powers and products of these terms. As such, the RHS of (\ref{maintransode}) will have an $n$-th Taylor coefficient that can be expressed as a recursive expression involving the first $n+1$ Taylor coefficients of $g$. Meanwhile, the LHS will give the $n+2$ Taylor coefficient of $g$ (up to a factor of $(n+1)(n+2)$), allowing one to recursive compute $\hat{A}_{n+2}$ from the previous Taylor coefficients $\{\hat{A}_0, \hat{A}_1, ..., \hat{A}_{n+1}\}$. More explicitly, we note that by the binomial theorem we have 
\begin{equation}
    \frac{S^2}{(1-\eta)^4} = S^2\sum_{n=0}^\infty \binom{n+3}{3} \eta^n.
\end{equation}
Hence by Cauchy's product rule and direct differentiation if we write 
\begin{equation}
    \frac{\displaystyle 2(1-\eta)^3 }{\displaystyle S^2}g' + \frac{\displaystyle  \left(\frac{\eta S}{1-\eta} + 1 \right)\left(\frac{(1-\eta)^4}{S^2}g'^2+1\right)^{3/2}Bg - \frac{(1-\eta)^6}{S^3}g'^3-\frac{(1-\eta)^2}{S}g'}{\displaystyle \frac{\eta S}{1-\eta}+1 } = \sum_{n=0}^\infty w_n \eta^n,
    \label{D4}
\end{equation}
then the $w_n$ will have a recursive expression in terms of $\hat{A}_n$ only up to $\hat{A}_{n+1}$, and equation (\ref{maintransode}) becomes
\begin{align} 
    \sum_{n=0}^\infty (n+1)(n+2) \hat{A}_{n+2} & = S^2\sum_{n=0}^\infty\left[ \sum_{k=0}^n \binom{k+3}{3}w_{n-k}\right]\eta^n,
    \\ \hat{A}_{n+2} &= \frac{S^2}{(n+1)(n+2)}\sum_{k=0}^n \binom{k+3}{3}w_{n-k}. \end{align} 
Continuing to work backwards, we can write by Cauchy's Product Rule that \begin{align}    
    w_n &= f_n + \sum_{k=0}^n (q_k - t_k - u_k)r_{n-k} 
\end{align} 
where 
\begin{align} 
    \frac{2(1-\eta)^3}{S^2} g' &= \sum_{n=0}^\infty f_n \eta^n 
    \\ \left(\frac{\eta S}{1-\eta}+1\right)\left( \frac{(1-\eta)^4}{S^2}g'^2+1\right)^{3/2}Bg &= \sum_{n=0}^\infty q_n \eta^n 
    \\ \frac{(1-\eta)^6}{S^3}g'^3 &= \sum_{n=0}^\infty t_n \eta^n \label{D10}
    \\ \frac{(1-\eta)^2}{S}g' &= \sum_{n=0}^\infty u_n\eta^n \label{D11}  
    \\ \left(\frac{\eta S}{1-\eta} + 1 \right)^{-1} &=\sum_{n=0}^\infty r_n \eta^n. \end{align}
We notice first that by term-by-term differentiation, Cauchy's product rule, and the binomial formula we have the following formulae for $f_n$ and $u_n$: 
\begin{align} 
    f_n &= \frac{2}{S^2}\sum_{k=0}^n (-1)^k \binom{3}{k}(n-k+1)\hat{A}_{n-k+1}, 
    \\ u_n &= \frac{1}{S}\sum_{k=0}^n (-1)^k \binom{2}{k} (n-k+1)\hat{A}_{n-k+1}. \end{align} 
Furthermore, by the formula for a geometric series, we have directly that 
\begin{align} 
    \left(\frac{\eta S}{1-\eta} + 1 \right)^{-1} &= \frac{(1-\eta)}{1-(1-S)\eta} 
    \\ r_n &= (1-S)^n - (1-\delta_{n,0})(1-S)^{n-1}.  
\end{align} 
where the Kronecker notation is used such that $\delta_{n,0}$ is 0 when $n\neq0$ and 1 when $n=0$. The computation of $t_n$ in~(\ref{D10}) could be obtained directly using JCP Miller's formula on $g'^3$ followed by Cauchy's product rule; instead, we first build the formula for the term involving $g'^2$ in~(\ref{D4}), and then  extract the result for $t_n$ from that. That is, we define 
\begin{equation} 
    \frac{(1-\eta)^4}{S^2}g'^2 = \sum_{n=0}^\infty d_n \eta^n,
    \label{D17}
\end{equation} 
and by Cauchy's product rule have 
\begin{align} 
    d_n &= \sum_{k=0}^n u_k u_{n-k}.
    \end{align}
The expression for $t_n$ can be obtained by taking the Cauchy product of~(\ref{D17}) and~(\ref{D11}) to yield    
\begin{align}
    t_n &= \sum_{k=0}^n d_k u_{n-k}. 
\end{align} 
From here it remains only for us to compute $q_n$. We begin by writing \begin{equation} 
    \frac{(1-\eta)^4}{S^2}g'^2 + 1 = \sum_{n=0}^\infty \bar{d}_n \eta^n,
\end{equation} 
then we have 
\begin{equation} 
    \bar{d}_n = d_n + \delta_{n,0}.
\end{equation} 
Proceeding in the same fashion if we write 
\begin{equation}
    \left(\frac{(1-\eta)^4}{S^2}g'^2 + 1\right)^{3/2} = \sum_{n=0}^\infty p_n \eta^n,
\end{equation} 
then by JCP Miller's formula we have 
\begin{equation} 
    p_{0} = \bar{d}_0^{3/2} \hphantom{\sum\sum} p_{n>1} = \frac{1}{n\bar{d}_0}\sum_{k=1}^n\left(\frac{5}{2}k-n\right)\bar{d}_kp_{n-k}.
\end{equation} 
Writing via geometric series that 
\begin{align*} 
    \frac{\eta S}{1-\eta}+1 &= \sum_{n=0}^\infty \bar{\ell}_n \eta^n 
    \\ \bar{\ell}_n &= S + \delta_{n,0} (1-S),
\end{align*} 
then, if we write 
\begin{equation} 
    \left(\frac{\eta S}{1-\eta}+1\right)g = \sum_{n=0}^\infty \ell_n \eta^n,  \end{equation} 
Cauchy's product rule gives 
\begin{equation}
    \ell_n = \sum_{k=0}^n \bar{\ell}_k \hat{A}_{n-k}. 
\end{equation} 
One final application of Cauchy's Product Rule completes the recurrence by writing 
\begin{equation} 
    q_n = B\sum_{k=0}^n \ell_k p_{n-k}. 
\end{equation}
In summary, the preceding coefficients $f_n$, $q_n$, $t_n$, $u_n$, $\ell_n$, $d_n$, $r_n$, and $p_n$ are all utilized in the power series solution~(\ref{eq:ConvPowerSeries_ODETransformed}) in the main text.

\section{Computation of the Euler coefficients in~(\ref{eq:ConvPowerSeries_ODETransformed_prefactor}) after inclusion of the modified Bessel function prefactor} \label{sec:Appendix_EuelerizedBessel}
Here, we re-sum to accelerate the convergence of the power series solution, making direct use of the previously derived series~(\ref{eq:ConvPowerSeries_ODETransformed}) written in terms of the Euler transformed variable.  The re-summation is enabled because the Euler coefficients in~(\ref{eq:ConvPowerSeries_ODETransformed}) are computationally stable.  This is in direct contrast to the original attempt to use the divergent series coefficients~(\ref{eq:DivPowerSeries}) to construct the Euler coefficients using~(\ref{eq:binomialEuler})--which leads to computational errors.   Here, we rewrite for reference~(\ref{eq:gseries}) as
\begin{subequations}
\begin{equation} g(\eta) = \sum_{n=0}^\infty \hat{A}_n \eta^n, \end{equation} where the coefficients $\hat{A}_n$ are given by~(\ref{eq:ConvPowerSeries_ODETransformed}) and 
\begin{equation}
    \eta(\tilde{r}) = \frac{\tilde{r}}{\tilde{r}+S}. 
    \label{eq:etatrans}
\end{equation}
\label{eq:gAhat}
\end{subequations}

We accelerate convergence of this summation by the inclusion of the leading asymptotic behavior in terms of the modified Bessel function of the second kind of zeroth order in accordance with~(\ref{eq:prefactor}) and~(\ref{eq:BesselEuler_Initial}), written for reference here as \begin{equation} \bar{h}(\tilde{r}) = K_0\left(\sqrt{B}(\tilde{r}+1)\right) \sum_{n=0}^\infty \bar{A}_n \left(\frac{\tilde{r}}{\tilde{r}+S}\right)^n.
\label{eq:resumBessel}
\end{equation} By direct rearrangement of~(\ref{eq:etatrans}), we have $\tilde{r} = \frac{S\eta}{1-\eta}$ , and noting that $h(r)=h(r(\eta))=g(\eta)$, we have 
\begin{subequations}
\begin{equation} g(\eta) = K_0\left(\sqrt{B}\left(\frac{\eta S}{1-\eta} + 1\right) \right) \sum_{n=0}^\infty \bar{A}_n \eta^n, \end{equation} 
or equivalently:
\begin{equation}
   g(\eta)\left[K_0\left(\sqrt{B}\left(\frac{\eta S}{1-\eta} + 1\right) \right)\right]^{-1} =  \sum_{n=0}^\infty \bar{A}_n \eta^n.
   \label{eq:prefactrecip}
\end{equation}
\label{eq:prefact}
\end{subequations}
Since $g(\eta)$ is known in terms of $\hat{A}_n$ from~(\ref{eq:gAhat}), we can obtained a relationship between $\bar{A}_n$ (the expansion with the prefactor) and $\hat{A}_n$ (the original Euler expansion) as follows.  We first define the argument of the Bessel function solution in~(\ref{eq:prefactor}) as
\begin{equation} 
\sqrt{B}(\tilde{r}+1)=\sqrt{B}\left(\frac{\eta S}{1-\eta} + 1\right)\equiv y(\eta),
\label{eq:ourf} 
\end{equation} 
and express the modified Bessel function portion of~(\ref{eq:prefact}) as
\begin{equation} K_0(y)=K_0\left(y(\eta)\right)=f(\eta)= \sum_{n=0}^\infty \bar{\zeta}_n \eta^n. 
\label{eq:BesselCoeffs}
\end{equation}
Using this expansion form, we employ JCP Miller's formula~(\ref{eq:JCP}) to yield:
\begin{subequations}
    \begin{equation}
        \left[K_0\left(\sqrt{B}\left(\frac{\eta S}{1-\eta} + 1\right)\right)\right]^{-1}=\sum_{n=0}^\infty \zeta_n\eta^n,
        \label{eq:recip}
    \end{equation}
    \begin{equation} \zeta_n = \frac{-1}{\bar{\zeta}_0} \sum_{k=1}^n \bar{\zeta}_k \zeta_{n-k} \hphantom{\sum\sum} \zeta_0 = \bar{\zeta}_0^{-1}. \label{eq:zetatrans}\end{equation}
\end{subequations}
Subsequently, by inserting~(\ref{eq:recip}) into~(\ref{eq:prefactrecip}) and using Cauchy's product rule~(\ref{eq:Cauchy}), we obtain \begin{equation} \bar{A}_n = \sum_{k=0}^n \zeta_k \hat{A}_{n-k}. 
\label{eq:recipCauchy}
\end{equation} The result~(\ref{eq:recipCauchy}) provides the necessary coefficients for the resummation--at this point, however, we have not yet determined the coefficients $\bar{\zeta}_n$ in~(\ref{eq:BesselCoeffs}), which we now obtain.

To proceed, we begin with the modified Bessel function governing equation~(\ref{eq:AsympODE1}) in the main text, rewritten in terms of $y$ as
\begin{equation} \frac{d^2K_0}{dy^2}  + \frac{1}{y}\frac{dK_0}{dy} - K_0 = 0. \label{eq:besseltrans} 
\end{equation} 
Our goal is to determine the Taylor expansion of $K_0(y(\eta))$ about $\eta=0$, which is precisely the expression~(\ref{eq:BesselCoeffs}) that we require. Employing the chain rule formulae given by~(\ref{eq:inversion}) (letting $F=K_0$), substituting this result into~(\ref{eq:besseltrans}), and solving for $\frac{d^2 f}{d\eta^2}$ leads to the expression
\begin{equation} 
\frac{d^2f}{d\eta^2} = \left(\frac[5pt]{\frac{d^2y}{d\eta^2}}{\frac{dy}{d\eta}
}~-~\frac[5pt]{\frac{dy}{d\eta}}{y}\right)\frac{d f}{d\eta} + \left(\frac{dy}{d\eta}\right)^2 f.
\label{eq:BesselEta}
\end{equation} 
 From here we will want to find the power series solution to~(\ref{eq:BesselEta}) in the form~(\ref{eq:BesselCoeffs}). For the particular $y$ given by~(\ref{eq:ourf}), we have the algebraic identities \begin{align}  \frac[5pt]{\frac{dy}{d\eta}}{y} &= \frac{1}{1-\eta}- \frac{1-S}{1-(1-S)\eta}, \\ \frac[5pt]{\frac{d^2y}{d\eta^2}}{\frac{dy}{d\eta}
} &= \frac{2}{1-\eta}, \\ \left(\frac{dy}{d\eta}\right)^2 &= \frac{BS^2}{(1-\eta)^4}. \end{align} Using the Binomial theorem, all of these expressions can be Taylor expanded explicitly as \begin{align} \label{eq:besseleulerizationderivationcoeffsstart} \frac[5pt]{\frac{dy}{d\eta}}{y} &= \frac{1}{1-\eta}- \frac{1-S}{1-(1-S)\eta} = \sum_{n=0}^\infty \rho_n \eta^n = \sum_{n=0}^\infty (1-(1-S)^{n+1}) \eta^n,~\rho_n = 1 - (1-S)^{n+1}, \\ \frac[5pt]{\frac{d^2y}{d\eta^2}}{\frac{dy}{d\eta}
} &= \frac{2}{1-\eta} = 2\sum_{n=0}^\infty \eta^n, \\ \left(\frac{dy}{d\eta}\right)^2 &= \frac{BS^2}{(1-\eta)^4} = BS^2 \sum_{n=0}^\infty  \binom{n+3}{3} \eta^n. \end{align} Furthermore, term-by-term differentiation of~(\ref{eq:BesselCoeffs}) gives \begin{align} \frac{df}{d\eta} &= \sum_{n=0}^\infty (n+1)\bar{\zeta}_{n+1} \eta^n, \\ \frac{d^2f}{d\eta^2} &= \sum_{n=0}^\infty (n+1)(n+2)\bar{\zeta}_{n+2} \eta^n. \label{eq:besseleulerizationderivationcoeffsend}  \end{align}
We substitute the expansion~(\ref{eq:BesselCoeffs}) and results~(\ref{eq:besseleulerizationderivationcoeffsstart})-(\ref{eq:besseleulerizationderivationcoeffsend}) into the ODE~(\ref{eq:BesselEta}), and after making use of Cauchy's product rule, we obtain
\begin{equation} \sum_{n=0}^\infty (n+1)(n+2) \bar{\zeta}_{n+2} \eta^n = \sum_{n=0}^\infty \left[\sum_{k=0}^n (k+1)\bar{\zeta}_{k+1}(2-\rho_{n-k})\right]\eta^n   + BS^2\sum_{n=0}^\infty \left[\sum_{k=0}^n \binom{k+3}{3}\bar{\zeta}_{n-k}\right]\eta^n .   \end{equation} Comparing like terms gives the recurrence relation
 \begin{subequations}
 \begin{equation} \bar{\zeta}_{n+2} = \frac{1}{(n+1)(n+2)}\sum_{k=0}^n \left\{ (k+1)\bar{\zeta}_{k+1}(2-\rho_{n-k}) + \binom{k+3}{3}\bar{\zeta}_{n-k}  \right\},  \end{equation} where we have the initial conditions by direct substitution that \begin{equation} \bar{\zeta}_0 = K_0\left( \sqrt{B}\right)~~\text{ and }~~\bar{\zeta}_1 = -S\sqrt{B} K_1\left(\sqrt{B}\right). \end{equation}
 \label{eq:end}
  \end{subequations}
The result~(\ref{eq:end}) can be inserted into~(\ref{eq:zetatrans}) and~(\ref{eq:recipCauchy}) which completely defines $\bar{A}_n$ in~(\ref{eq:resumBessel}) and~(\ref{eq:BesselEuler_Initial}), and the resummation is thus complete.

\bibliographystyle{imamat}

\bibliography{CylindricalWallBib}
\end{document}